\documentclass[a4paper,11pt]{article}
\usepackage{epsfig,amsmath,amsfonts,breqn}
\usepackage{appendix}
\usepackage[T1]{fontenc}
\usepackage{dsfont}
\usepackage[latin1]{inputenc}  

\usepackage{amsthm}

\usepackage{color}
\usepackage{paralist}
\usepackage{graphicx,psfrag}

\usepackage{tikz}

\usepackage{scalefnt}

\usepackage{rotating}
\usepackage{fix-cm}

\usepackage[ruled,vlined]{algorithm2e}

\usepackage[misc,geometry]{ifsym}

\def\XXint#1#2#3{{\setbox0=\hbox{$#1{#2#3}{\int}$ }
		\vcenter{\hbox{$#2#3$ }}\kern-.6\wd0}}

\theoremstyle{definition}
\newtheorem{definition}{Definition}

\theoremstyle{plain}
\theoremstyle{plain}

\theoremstyle{plain}

\theoremstyle{plain}

\newcommand{\R}{\mathds{R}}

\newcommand{\C}{\mathds{C}}

\begin{document}
	\begin{center}
		{\LARGE \bf Study on the Behavior of Weakly Nonlinear Water Waves in the Presence of Random Wind Forcing} \\[3mm]
		{\large L. Dostal$^1$, M. Hollm$^{1}$ and  E. Kreuzer$^{1}$}\\[2mm]
		$^1$\small Institute of Mechanics and Ocean Engineering, Hamburg University of Technology, Hamburg, Germany\label{firstpage}
	\end{center}

	\begin{abstract}
		Specific solutions of the nonlinear Schr\"odinger equation, such as the Peregrine breather, are considered to be prototypes of extreme or freak waves in the oceans. An important question is, whether these solutions also exist in the presence of gusty wind. 
		Using the method of multiple scales, a nonlinear Schr\"odinger equation is obtained for the case of wind forced weakly nonlinear deep water waves. Thereby, the wind forcing is modeled as a stochastic process.
		This leads to a stochastic nonlinear Schr\"odinger equation,
		which is calculated for different wind regimes.
		For the case of wind forcing which is either random in time or random in space, it is shown that breather type solutions such as the Peregrine breather occur even in strong gusty wind conditions.
	\end{abstract}
	\begin{center}
	\end{center}

\section{Introduction}\label{sec:Introduction}
%
\textcolor{black}{Extreme or freak waves endanger the life of offshore workers, crews, and passengers of ships, and can cause major damage to offshore structures and vessels. Such very large waves were measured in the oceans, as was for example reported in \cite{Kharif:2009}. Famous measured waves are the Draupner wave and the Yura waves. 
	%
	Up to now, it is not well understood how high and extreme ocean waves arise. For the design and determination of operational limits of offshore structures and floating bodies in the sea environment, the physically {appropriate} modeling of sea states is a task of great importance. A closer look at the sea surface immediately reveals the irregularity of ocean gravity waves. This irregularity is caused by many influences, such as wind, swell, and currents. These influences can in general be quantified only statistically.}
%
%
%
%
%
%
%
 \textcolor{black}{Therefore, accurate modeling of sea states is based on random fields and known as random seas. For simulations of linear random seas, it is common to superimpose harmonic waves with random phase shifts, whereby the amplitudes and angular frequencies of these harmonic waves are obtained from the spectrum of the desired sea conditions. Because a finite number of harmonic functions is superimposed in this procedure, the resulting sea state is periodic. As the period increases with the number of harmonic wave components, also the necessary computation time for random wave generation increases.
 	A better way for simulations is to generate the stochastic process by a continuous autoregressive moving average (CARMA) process, having the same spectral density \cite{brockwell:1995,dostal:2011}. Then the random wave field is created efficiently by integrating a system of stochastic differential equations, where the resulting random wave field is not periodic.}
 %
 %
 %
 
 \textcolor{black}{Satellite observations revealed that extreme waves appear more often then predicted by linear wave theory, as was reported in results of the MaxWave project \cite{rosenthal:2008}.
 	In a sea state determined by superposition of linear waves (Airy waves), the distribution of wave elevation is Gaussian and the wave amplitudes are Rayleigh distributed.
 	However, experimental results at Marintek \cite{onorato:2004} show that, in sea states with a Benjamin-Fair index (BFI) of about one or higher, the Rayleigh distribution clearly underestimates the tail of the probability density function of wave heights.
 	This implies that linear random wave theory is not sufficient for determination of the distribution of high and extreme waves. Therefore, it is also not sufficient to determine safety limits for ships and offshore structures only according to linear random wave theory. }
 
 \textcolor{black}{Several theories were developed to account for the deviation of wave heights from the Rayleigh distribution and take into account the nonlinear behavior of ocean waves. 
 	Nonlinear modeling of random seas allows for the analysis of waves with a significantly greater wave slope then it is possible with linear wave theory. 
 	Second-order perturbation expansion of randomly superposed waves already well describes most rogue wave statistics  \cite{dysthe2008oceanic}. 
 	The second-order random seas superposition solution can be found in \cite{sharma1981second,forristall2000wave,agarwal2011incorporating}. 
 	This solution includes bound harmonics, which are phase locked to corresponding linear wave components \cite{gibson2007evolution}.   
 	The deviation of a random wave field from a Gaussian wave field by calculating the bound harmonics non-Gaussianity is determined in \cite{janssen2009some,annenkov2013large}.
 	For narrow band second-order random seas the probability density function was obtained by Tayfun \cite{tayfun1980narrow}. }
 
 \textcolor{black}{Waves that meet the rogue wave definition of being at least two times higher than the significant wave height can be obtained by various different mechanisms. Such mechanisms are geometric focusing, dispersive focusing and nonlinear focusing \cite{dysthe2008oceanic}. Nonlinear focusing appears in \textcolor{black}{long-crested} sea states, as was shown by Gramstad and Trulsen \cite{gramstad2007influence}.}

 \textcolor{black}{It is sufficient for many problems in offshore engineering to solve the Euler equations, instead of considering the Navier-Stokes equations. \textcolor{black}{But these equations are still very expensive to solve numerically, since one needs an approach which can deal with the flexibility of the structure and the nonlinear and stochastic nature of the waves and the wind \cite{marino2011fully}. Although there exist some algorithms to compute this, which are for example presented in \cite{dold1986efficient,longuet1976deformation,dommermuth1988deep,grilli1989efficient}, they have all in common that especially the long-time simulations are very time consuming.}
 	%
 	%
 	%
 	%
 	It can be shown that weakly nonlinear solutions of the Euler equations can be reduced to solutions described by complex envelopes, which satisfy the nonlinear Schr\"odinger equation (NLS), as was derived by Zakharov \cite{zakharov:1968}. Such reduction can be achieved by the method of multiple scales, cf. \cite{Mei:1983}.
 	The envelope of nonlinear waves, such as water waves or optical waves, can then be described by the NLS. \textcolor{black}{Although the solutions of the NLS only approximate the solutions of the Euler equations, it is much easier to find them, whereby the nonlinear and stochastic nature of the wind and waves is considered as well. Besides this, in the one dimensional case a variety of solutions of the deterministic NLS} have been presented in
 	\cite{kuznetsov1977solitons,akhmediev1985generation,peregrine1983water}. These solutions are known as Akhmediev, Kuznetsov-Ma, and Peregrine breathers. The Peregrine breather is localized \textcolor{black}{in time and space}, since it is the limiting case of the Akhmediev and Kuznetsov-Ma breathers. Therefore, this solution describes a wave, which seems to come from nowhere and disappears without a trace \cite{akhmediev2009waves}. Such a behavior is known from freak waves, as measured for example at the Draupner platform \cite{Kharif:2009}.
 	It was experimentally shown that the so called breather solutions of the NLS can be generated in a wave flume \cite{chabchoub:2011}.
 	Experimental results by Chabchoub~\cite{chabchoub2016tracking} show observations of Peregrine breathers in random sea states. Therefore, a further studies of the NLS under random perturbations are needed for new insights  about nonlinear stochastic sea states and the behavior of nonlinear random waves.
    This will allow to analyze whether it makes sense to use loads due to Peregrine solutions for sea-keeping tests of ships and offshore structures, which were carried out for example in~\cite{onorato2013rogue,klein2016peregrine}.}
The main growth of gravity waves is induced by wind. \textcolor{black}{In order to model the influece of wind on water waves, three mechanisms can be considered: the Jeffrey' sheltering \cite{jeffreys1925formation}, the Phillips \cite{phillips1957generation} and the Miles' mechanism \cite{miles:1957,miles:1959}. While Jeffrey assumed that the growth of steep waves comes from the separation of the air flow on the lee side of the wave crest, the theories of Phillips and Miles state that the waves were generated by a resonance phenomenon \cite{montalvo2013wind}. Hereby, Phillips considered resonance between the surface waves and turbulent pressure fluctuation in the air, while Miles considered resonance beteween the waves and the wave-induced pressure fluctuations. The mechanism of Jeffrey is well suited for shallow water waves, which are in general steeper than deep water waves \cite{chambarel2010generation}. Moreover, Phillips' mechanism turned out to be ineffective, since independent of the wave spectrum, the effect is of the order of the square of the density ratio of air and water. The Miles' mechanism is therefore a well-established model for the forcing of weakly nonlinear water waves \cite{janssen2004interaction}, whereby a quasi-laminar approximation is used leading to a stream function, which satisfies the Rayleigh equation. It is of the order of the air-water density ratio and, on the other hand, the practical relevance of Miles mechanism has been confirmed in numerical simulations \cite{alexakis2004nonlinear} and in field experiments for long waves \cite{hristov2003dynamical} in deep water.} 

 	Using the method of multiple scales, Leblanc~\cite{leblanc2007amplification} derived a forced NLS for the case of deterministic wind forcing for weakly nonlinear surface gravity waves. Later, deterministic wind forcing and viscous dissipation for weakly nonlinear surface gravity waves by means of a forced and damped NLS was considered by Kharif~et~al.~\cite{Kharif:2010}. {Higher order approximation of the nonlinear water wave envelope leads to Dysthe equations \cite{dysthe:1979}, for which fluid viscosity has been included in \cite{carter2016frequency}.}
 
 \textcolor{black}{Experimental studies on the effect of wind forcing on the modulation instability and the Peregrine breather were presented in \cite{chabchoub2013experiments}, whereby the Peregrine breather has been also detected under strong wind conditions. Results concerning the effect of strong time-invariant wind on the modulation instability were obtained in \cite{brunetti2014nonlinear,brunetti2014modulational}. Moreover, the spectral up- and downshifting of Akhmediev breathers under wind forcing has been shown recently numerically and experimentally \cite{eeltink2017spectral} using higher order modification of the nonlinear Schr\"odinger equation and Miles mechanism \cite{miles:1957}.}
 %
 %
 %

 \textcolor{black}{Combining the NLS with a random excitation process leads to a stochastic partial differential equation (SPDE).
 	Since the theory on stochastic partial differential equations is a vast field, only a few related references {are mentioned.} 
 	%
 	In \cite{Antoine2013,rapti2004c,rapti2003} the nonlinear Schr\"odinger/Gross-Pitaevskii equation for modeling Bose-Einstein condensation and nonlinear optics was treated by different numerical methods including finite difference time domain methods, the relaxation method, and the time-splitting spectral method. An analysis of wave motion under stochastic excitation is presented in \cite{Namachchivaya2010}. Parametric and modulation instabilities arising in a
 	non-autonomous, discrete NLS were analyzed by Rapti et al. \cite{rapti2004b}.
 	For the stochastic NLS in optics and Bose-Einstein condensation, finite difference schemes were derived by Debussche, de Bouard and Di Menza in \cite{Bouard1999,DeBouard2001,Debussche2002,DeBouard2006} in order to obtain numerical solutions.
 	Moreover, in \cite{Debussche2002,Bouard2002} additive or multiplicative real space-time white noise is considered.}

 In this work, weakly nonlinear water waves are modeled by the NLS and their behavior in the presence of random wind forcing is studied with focus on solitons and breather solutions. 

 {This work is organized as follows.} In section~\ref{sec:wind_wave_models} the NLS for nonlinear dissipative waves excited by wind is introduced. Deterministic soliton and Peregrine solutions of the NLS are shown in section~\ref{sec:deterministic NLS}, followed by a short discussion and validation of the numerical solution method used for the NLS in section~\ref{sec:numerical solution method}. Then results on the NLS excited by strong white noise {are presented} in section~\ref{sec:white noise excitation}, before modeling random wind processes and calculating the NLS excited by random wind in sections~\ref{sec:RandomWind} and \ref{sec:solitonPeregrineRandomWind}, respectively.

\section{Weakly nonlinear dissipative waves excited by wind}\label{sec:wind_wave_models}
It can be shown that weakly nonlinear solutions of the Euler equations can be reduced to solutions described by a complex envelope, which satisfies the NLS. Such reduction can be achieved by the method of multiple scales, cf. \cite{Mei:1983}.
Following \cite{leblanc2007amplification,Leblanc2008,Kharif:2010}, the physical modeling of damped narrow-banded weakly nonlinear surface gravity waves, which are excited by steady wind, leads to a perturbed NLS. 
%
%
%
%
%

\subsection{Forced nonlinear Schr\"odinger equation for time and space variant wind-induced pressure}
{An essential part of the presented research is the consideration of a time and space variant wind-induced pressure in the forced Euler equations, which are given in \textcolor{black}{Appendix~1}. 
	In order to rigorously derive a nonlinear Schr\"odinger equation from the forced Euler equations 
	for the case of weakly nonlinear water waves, the method of multiple scales is used, whereby terms up to the third order in wave steepness $\varepsilon$ are considered. This derivation is shown in \textcolor{black}{Appendix~1} and leads \textcolor{black}{ in deep water} to the following perturbed NLS for the case of deep water and nonzero viscosity 
	%
	\begin{equation}\label{eq:nlsphysical}
	\begin{aligned}
	\mathrm{i}\,\psi_{\tau}-\frac{\omega}{8 k^2}\,\psi_{\xi\xi}-\frac{1}{2}\omega k^2\,|\psi|^2\psi=-\mathrm{i}\,k\,\frac { p^{1,1}}{\omega\rho_{w}}-2\,\mathrm{i}\,\,\nu\,k^2\,\psi,
	\end{aligned}
	\end{equation}
	where ${\xi=\varepsilon(x-c_gt)}$ is a spatial coordinate moving with the deep water group velocity $c_g=\frac{\omega}{2k}$, $\tau=\varepsilon^2\,t$ is the scaled time, $\psi(\xi,\tau)\in \C$ is the wave envelope,  $\varepsilon \ll 1$ is the wave steepness, $\nu$ is the viscosity of water, $k$ is the wave number, $\omega$ is the frequency of a carrier wave and $p^{1,1}(\xi,\tau)$ is the leading component of the wind-induced pressure $P_a$, assuming that $P_a$ is of order $\mathcal{O}(\varepsilon^3)$.%
	%
}
%
%
%
%
\textcolor{black}{The free surface elevation $\eta$ of a weakly nonlinear dispersive gravity wave on deep water is obtained from the NLS \eqref{eq:nlsphysical} by 
	\begin{equation}\label{eq:surfaceelevation}
	\eta({x}, t) = \psi({x}, t)\exp\{\mathrm{i}\,(k\,{x}-\omega\,t)\} +c.c.,
	\end{equation}
	where $c.c.$ denotes the complex conjugate.
	Moreover, 
	the} wave period $T$ and the wave length $\lambda$ are explicitly given by
\begin{equation}
T=\frac{2\pi}{\omega},\quad \quad \lambda=\frac{g}{2\pi}T^2.
\end{equation}
%
%
%
%
%
%
\subsection{Miles mechanism}
The well-known Miles mechanism  has proven to be a simple, yet versatile model for wind-induced wave growth \cite{hristov2003dynamical}.
For deterministic surface elevation $\eta(x,t)$, Miles assumed in \cite{miles:1957}, that the aerodynamic pressure is given by ${P_a=(\alpha+\mathrm{i}\beta)\rho_a\,U_1^2 k\,\eta(x,t)}$, where $\alpha$ and $\beta$ are coefficients, $k$ is the wave number, $\rho_a$ is density of air and $U_1$ is a characteristic wind velocity. 

For a logarithmic velocity profile in the boundary layer the characteristic velocity is $U_1=u_*/\kappa$,
where $\kappa$ is the von Karman constant and $\beta$ was obtained by Miles in \cite{miles:1959,miles:1996}.

Using only the pressure component which is in phase with the wave slope, the aerodynamic pressure can be further simplified to \cite{Kharif:2010}
\begin{equation}\label{eq:pressure}
P_a= \rho_a\,\beta\,\left(\frac{u_*}{\kappa}\right)^2\eta_x(x,t).
\end{equation}
{In this study, a simple extension to the Miles mechanism is used, whereby the friction velocity $u_*$ is time- and space-variant in the expression \eqref{eq:pressure} for the wind-induced pressure $P_a$.
	Substituting the expansion \eqref{eq:eta_ansatz} of $\eta$ and the expansion \eqref{eq:p_ansatz} of $P_a$ into equation \eqref{eq:pressure} and collecting terms of same order in $\varepsilon$ yields for $p^{1,1}$
		\begin{equation}
		\begin{aligned}
		p^{1,1}(\xi,\tau) =-\frac{\omega^2}{2 g} \rho_a\,\beta\,\left(\frac{u_*(\xi,\tau)}{\kappa}\right)^2\psi(\xi,\tau).
		\end{aligned}
		\end{equation}
%
%
	Substituting this result into equation~\eqref{eq:nlsphysical}, the following forced NLS is obtained
	\begin{equation}\label{eq:nlsdeterministic}
	\mathrm{i}\,\psi_{\tau}-\frac{\omega}{8 k^2}\,\psi_{\xi\xi}-\frac{1}{2}\omega k^2\,|\psi|^2\psi=\mathrm{i}\Gamma(\xi,\tau)\psi,
	\end{equation}
	with
	\begin{equation}\label{eq:Gamma_neu}
	\Gamma(\xi,\tau)=\frac{k\omega}{2 g}\frac {\rho_a}{\rho_{w}} \,\beta\,\left(\frac{u_*(\xi,\tau)}{\kappa}\right)^2  -2\,\nu\,k^2.
	\end{equation}
	If the friction velocity in the forcing term $\Gamma$ is assumed to be a stochastic process, then
	${\zeta(\xi,\tau):=\Gamma(\xi,\tau)}$ is also a stochastic process and the following stochastic NLS is obtained
	\begin{equation}\label{eq:nlsnoise}
	\begin{aligned}
	\mathrm{i}\,\psi_{\tau}-\frac{\omega}{8k^2}\,\psi_{\xi\xi}-\frac{1}{2}\,\omega\,k^2\,|\psi|^2\psi=\,\mathrm{i}\,\zeta\,\psi.
	\end{aligned}
	\end{equation}
}

For a logarithmic wind profile the relation between wind velocity $U(\xi,z,\tau)$ at height $z$ and friction velocity $u_*(\xi,\tau)$ is given by
\begin{equation}\label{eq:logWindProfile}
U(\xi,z,\tau)=\frac{u_*(\xi,\tau)}{\kappa}\ln\left(\frac{z}{z_0}\right),
\end{equation}
where the roughness length is given by $z_0=\alpha_{ch}u_*^2/g$, with the Charnock constant ${\alpha_{ch}\approx 0.01875}$. \textcolor{black}{Considering equation \eqref{eq:logWindProfile}, the friction velocity $u_*(\xi,\tau)$ has to be computed iteratively for a given wind velocity $U(\xi,z,\tau)$ at a prescribed \textcolor{black}{height} $z$.}
%
%
For the values of $\beta$, which are a function of $\kappa c_0/u_*$, the results of Conte and Miles \cite{conte1959numerical} for the logarithmic wind profile~\eqref{eq:logWindProfile} are used. {Thereby, the dimensionless roughness length ${\kappa^2 g z_0/u_*^2=0.003}$ is chosen for numerical calculations.}

\section{Deterministic soliton and Peregrine solutions of the NLS}\label{sec:deterministic NLS} 
Results on deterministic soliton and breather solutions of the NLS are well known and closed-form expressions are available for the undisturbed case. Recently also a closed-form expression of the Peregrine breather for the case with constant forcing has been {obtained} by Onorato and Proment~\cite{onorato2012approximate}. 
\subsection{Soliton solution}
\begin{figure}[h]
	\centering	
	\includegraphics[width=1.0\columnwidth]{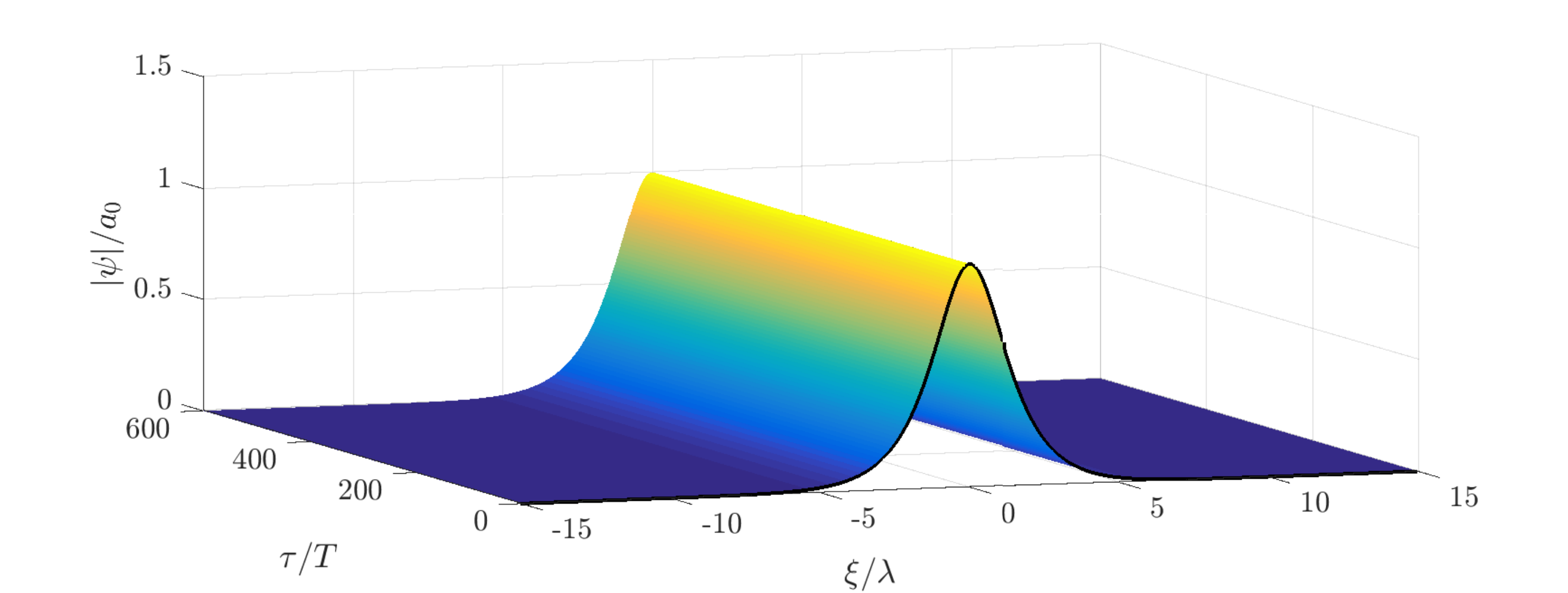}%
	\caption{Temporal evolution of a soliton solution of the stochastic NLS.}%
	\label{fig:soliton}%
\end{figure}

First we consider the unperturbed NLS, which is given by equation \eqref{eq:nlsdeterministic} with $\Gamma=0$ \textcolor{black}{1/s}.
For this case, a stationary soliton solution is given by
\begin{equation}\label{eq:solitonSolution}
\psi=a_0\,\mathrm{sech}(\sqrt{2}\,a_0 k^2 \,x)\exp(-\mathrm{i}\frac{1}{4}\,|a_0\,k|^2\,\omega\,t),
\end{equation}
with the free background amplitude parameter $a_0$. 
The corresponding soliton solution is shown in Fig.~\ref{fig:soliton}.

\subsection{Peregrine solution}
An important localized solution in time and space of the unperturbed NLS has been found by Peregrine \cite{peregrine1983water} and is known as the Peregrine breather. This solution is a rogue wave prototype, since its peak amplitude is several times higher than the amplitude of the background waves. For the NLS with constant forcing, a Peregrine breather solution was obtained by Onorato and Proment~\cite{onorato2012approximate}. This solution is given by
\begin{equation}\label{eq:forced breatherOnorato}
\psi=\exp(\Gamma t)a_0G(x,t)\left( \frac{ 4a(1-\mathrm{i}2ba_0^2p(t)t)}{a+a(2ba_0^2p(t)t)^2+2ba_0^2(p(t)x)^2} -1\right),
\end{equation}
where
\begin{equation}
G(x,t)=\sqrt{p(t)}\exp\left( \mathrm{i} \left( \frac{\Gamma p(t) x^2}{2a}-ba_0^2p(t)t \right)\right),
\end{equation}
and
\begin{equation}
a= \frac{\omega}{8 k^2}, \quad\quad\quad b=\frac{1}{2}\omega k^2.
\end{equation}
Thereby, the transformation $p(t)=1/(1-2\Gamma t)$ has been used, which has a singularity at ${|2\Gamma t|=1}$.

For $\Gamma=0$ \textcolor{black}{1/s}, equation \eqref{eq:forced breatherOnorato} reduces to the unperturbed Peregrine solution, which is given by
\begin{equation}\label{eq:unforced breatherOnorato}
\psi=a_0\exp\left(-\mathrm{i}ba_0^2t\right)\left( \frac{ 4a(1-\mathrm{i}2ba_0^2t)}{a+a(2ba_0^2t)^2+2ba_0^2x^2} -1\right).
\end{equation}
%

\section{Numerical solution method for the NLS}\label{sec:numerical solution method}
%
Different methods can be applied for the solutions of the various models of surface gravity waves.
%
%
For the calculation of solutions of nonlinear water wave surface elevation excited by a random wind process, stochastic partial differential equations have to be solved numerically. Moreover, the involved random process has to be non-white.
%

In order to obtain numerical solutions for the deterministic and stochastic NLS \eqref{eq:nlsdeterministic} and \eqref{eq:nlsnoise}, a relaxation finite difference scheme related to the schemes used by Debussche, de Bouard and di Menza in \cite{Bouard1999,DeBouard2001,Debussche2002,DeBouard2006} is used. The relaxation scheme has been introduced by Besse \cite{besse2004relaxation} as an extension to schemes of Crank-Nicolson type. In contrast to Crank-Nicolson type schemes, the relaxation scheme does not need to fulfill a Courant-Friedrichs-Lewy (CFL) condition, which links the discretization in time to the discretization in space and can make the numerical computation infeasible. The relaxation scheme is further discussed in \textcolor{black}{Appendix~2}. 

\begin{figure}[htb]
	\centering	
	\includegraphics[width=0.99\columnwidth]{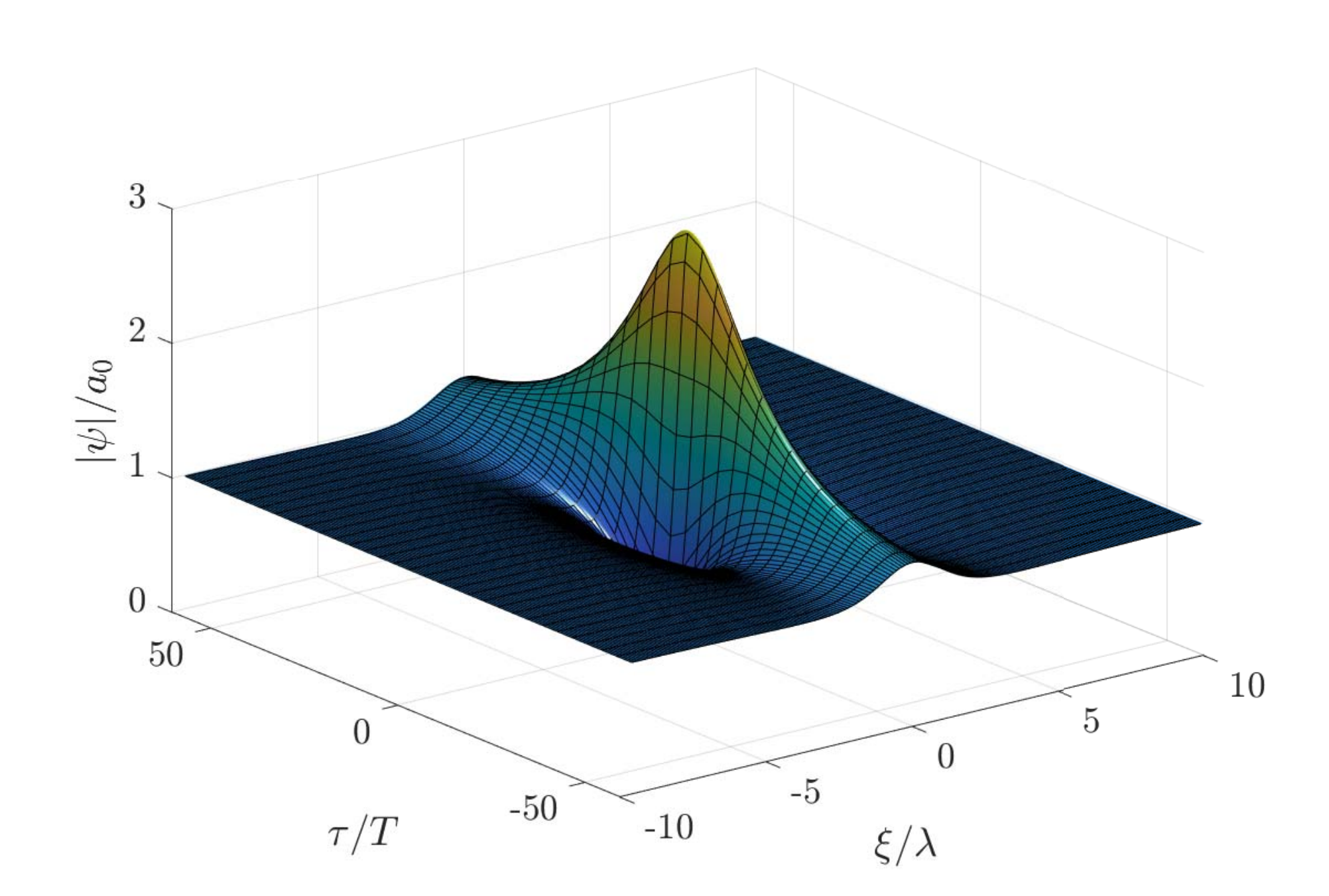}%
	\caption{Comparison of the numerical forced Peregrine solution (surface) with the analytical solution (mesh).}%
	\label{fig:PeregrineSolutionNumeric}%
\end{figure}

\begin{figure}[htb]
	\centering	
	\hspace{-0.3cm}
	\includegraphics[width=0.51\columnwidth]{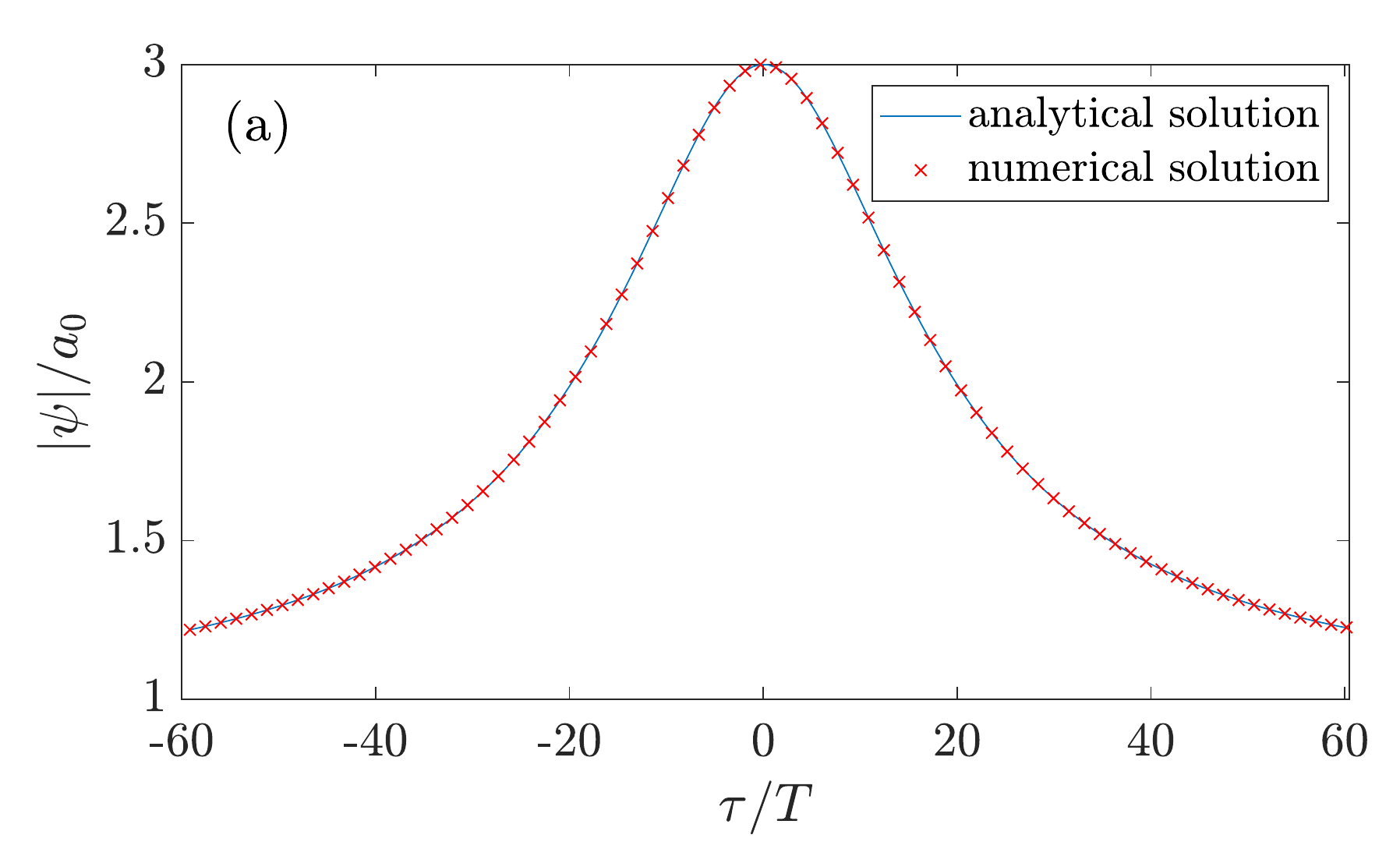}%
	\hspace{-0.2cm}
	\includegraphics[width=0.51\columnwidth]{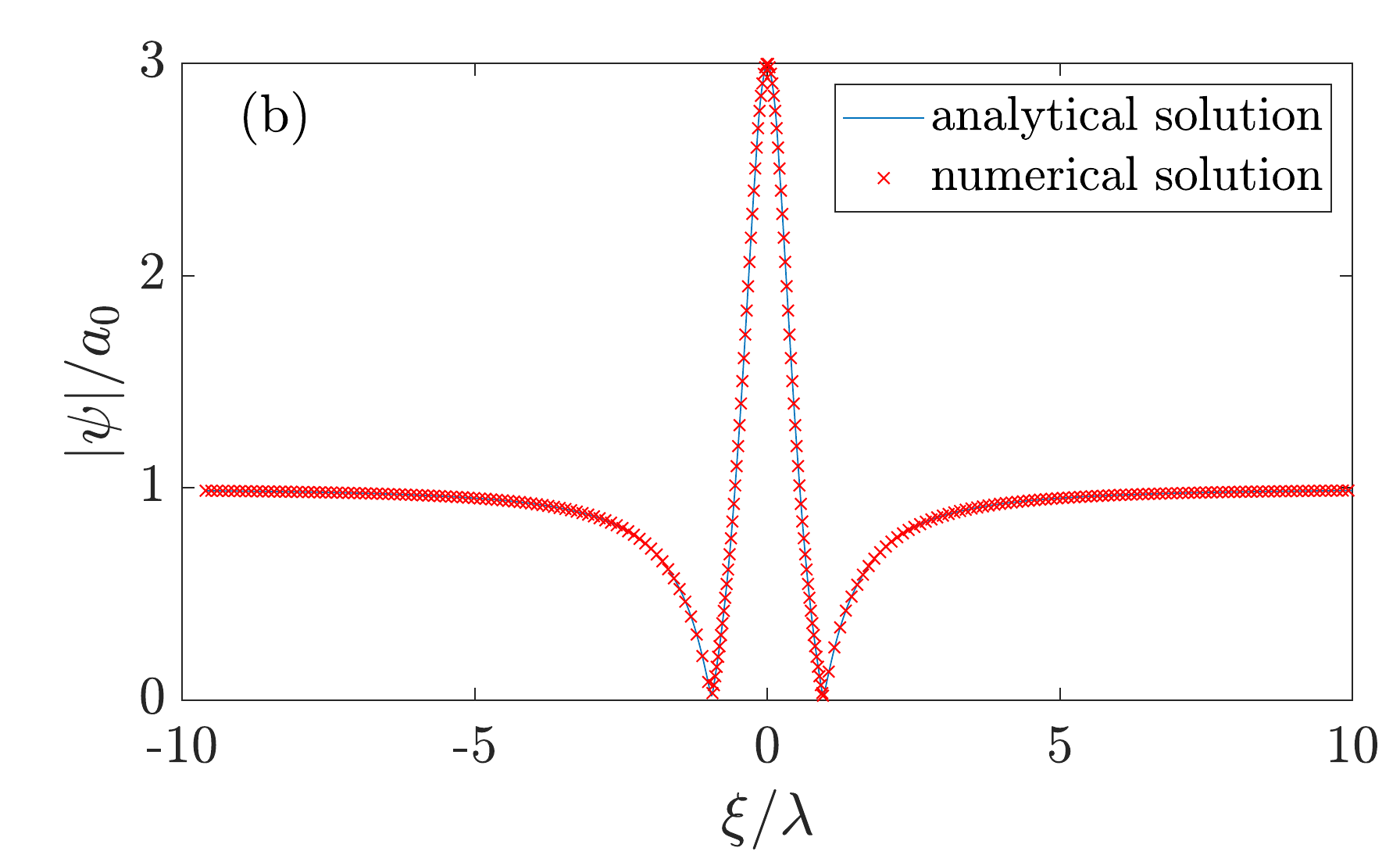}%
	\caption{Comparison of the time evolutions of numerical forced Peregrine solution with the analytical solution at (a) the fixed position $\xi/\lambda=-1.8$ and (b) the fixed timepoint $\tau/T=0$.}%
	\label{fig:PeregrineSolutionNumeric4}
\end{figure}

\begin{figure}[htb]
	\centering	                                      
	\includegraphics[width=0.8\columnwidth]{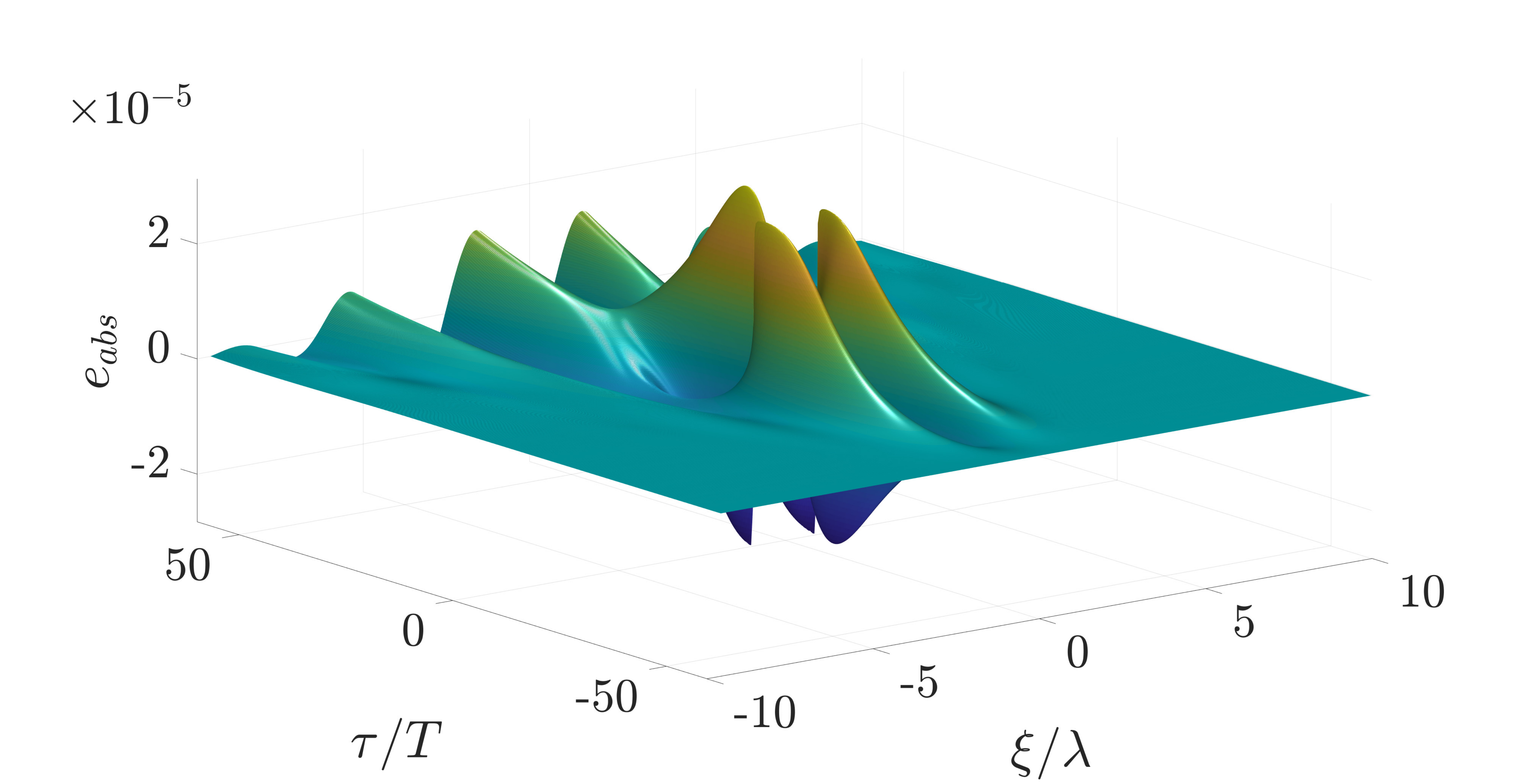}%
	\caption{Absolute error between the numerical forced Peregrine solution and the analytical solution.}%
	\label{fig:PeregrineSolution_absolute_error}%
\end{figure}

\begin{figure}[htb]
	\centering	                                      
	\includegraphics[width=0.8\columnwidth]{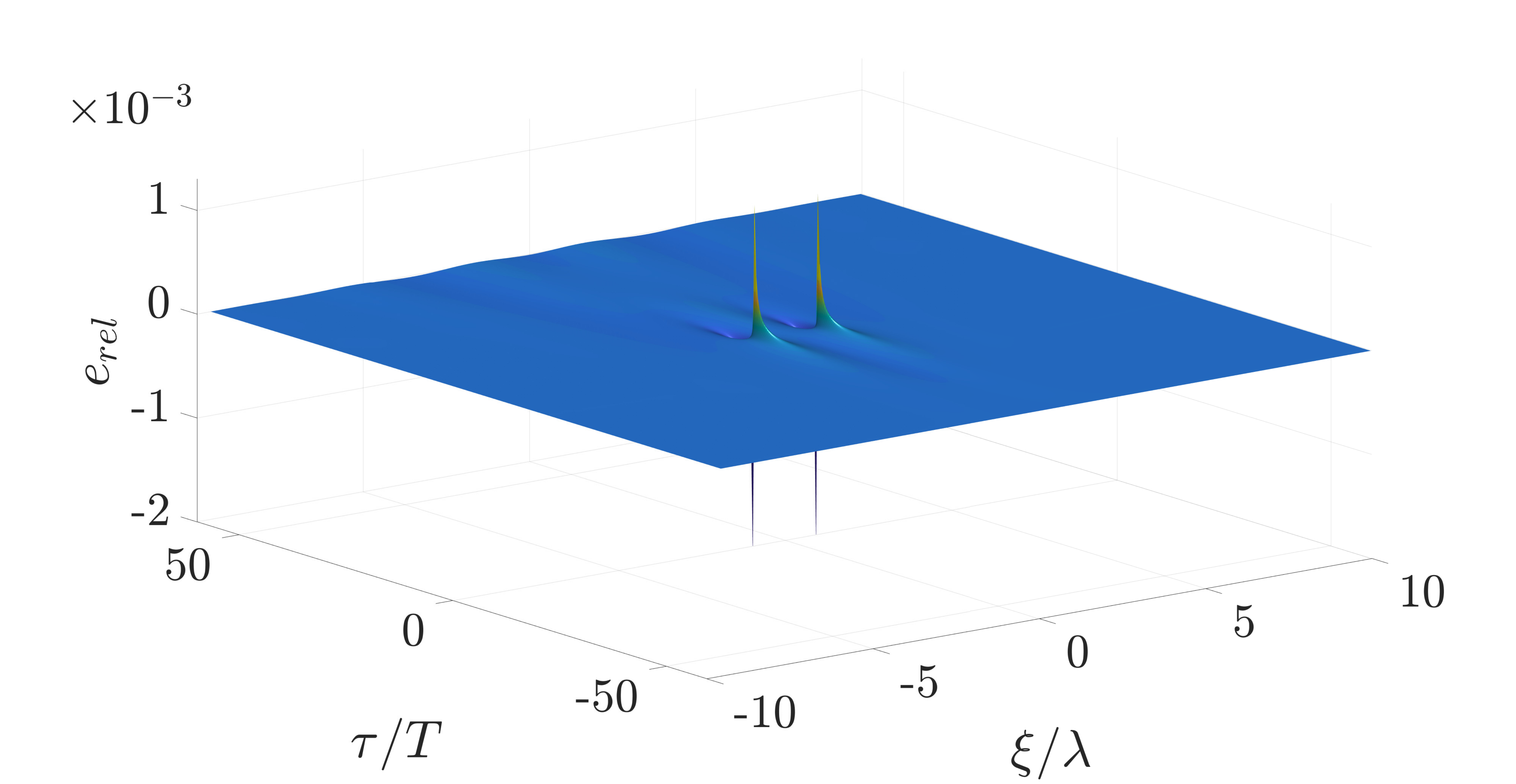}%
	\caption{Relative error between the numerical forced Peregrine solution and the analytical  solution.}%
	\label{fig:PeregrineSolution_reltive_error}%
\end{figure}

For the numerical solution of the deterministic and stochastic NLS \eqref{eq:nlsdeterministic} and \eqref{eq:nlsnoise}, periodic boundary conditions on a large enough domain $(x,t)\in D\subset\R \times \R^+$ {are considered, as well as} an initial condition $\psi(x,0)=\psi_0(x)$ at time $t=0$. 
Using the relaxation scheme, 
numerical results of the Peregrine breather solution {are obtained}, which are very close to the  corresponding analytical solution. 
In Fig.~\ref{fig:PeregrineSolutionNumeric}, a numerical solution of the NLS \eqref{eq:nlsdeterministic} together with the analytical solution from equation~\eqref{eq:forced breatherOnorato} are shown for the unexcited case\textcolor{black}{, i. e. $\Gamma=0$ }{1/s}. \textcolor{black}{The solutions for a fixed position and time point are shown in Fig.~\ref{fig:PeregrineSolutionNumeric4} and the} corresponding absolute and relative error is shown in Fig.~\ref{fig:PeregrineSolution_absolute_error} and Fig.~\ref{fig:PeregrineSolution_reltive_error}, respectively. As can be seen from these figures, the error is very small. Although the relative error includes two higher peaks, which result from the fact that the analytical solution has an amplitude near zero at these points, the relaxation scheme is very suitable for numerical calculations of the forced NLS.




\begin{figure}[htb]
	\centering	
	\includegraphics[width=1.05\columnwidth]{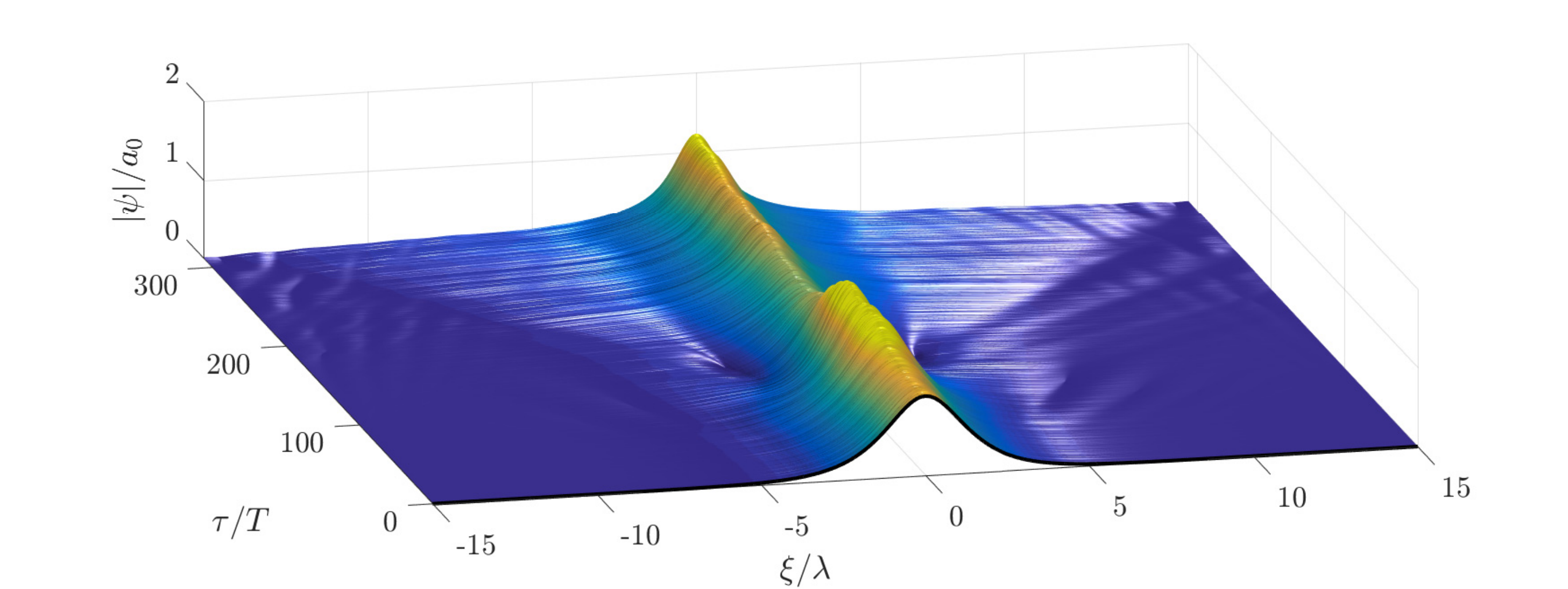}%
	\caption{Temporal evolution of a soliton solution of the stochastic NLS with white noise excitation in time, strong random forcing $\zeta_{\tau}$ with a variance ${\sigma^2= 0.2 \, \mathrm{1/s^2}}$.}%
	\label{fig:white noise_solitonSolutionnoblowup}%
\end{figure}


\section{Random excitation by white noise}\label{sec:white noise excitation}
At first a Gaussian white noise process $\zeta(\tau)$ in time $\tau \in \R$ is considered as the random forcing of the NLS \eqref{eq:nlsnoise}. This process has the properties ${E\{\zeta(\tau)\zeta(\tau+s)\}=\sigma^2\delta(s)}$,
${s,\,\sigma\in \R}$, where  $\delta(\cdot)$ is the Dirac function, and $E\{\zeta(\tau)\}=0$.


With this, sample results for the stochastic NLS \eqref{eq:nlsnoise} under white noise excitation are calculated. The random excitation due to such a white noise process reveals fundamental random dynamics of the stochastic NLS \eqref{eq:nlsnoise}.
%
A sample solution of a perturbed soliton with $a_0=1 \, \mathrm{m}$ and $\omega=1 \, \mathrm{rad/s}$, which is excited by white noise in time according to equation \eqref{eq:nlsnoise}, is shown in Fig.~\ref{fig:white noise_solitonSolutionnoblowup}. Thereby, the variance of the white noise excitation has been set to \textcolor{black}{${\sigma^2= 0.2 \, \mathrm{1/s^2}}$}, and the same initial condition as in the case of the unperturbed soliton from equation \eqref{eq:solitonSolution} has been chosen. 
%

\begin{figure}[htb]
	\centering	
	\includegraphics[width=1.05\columnwidth]{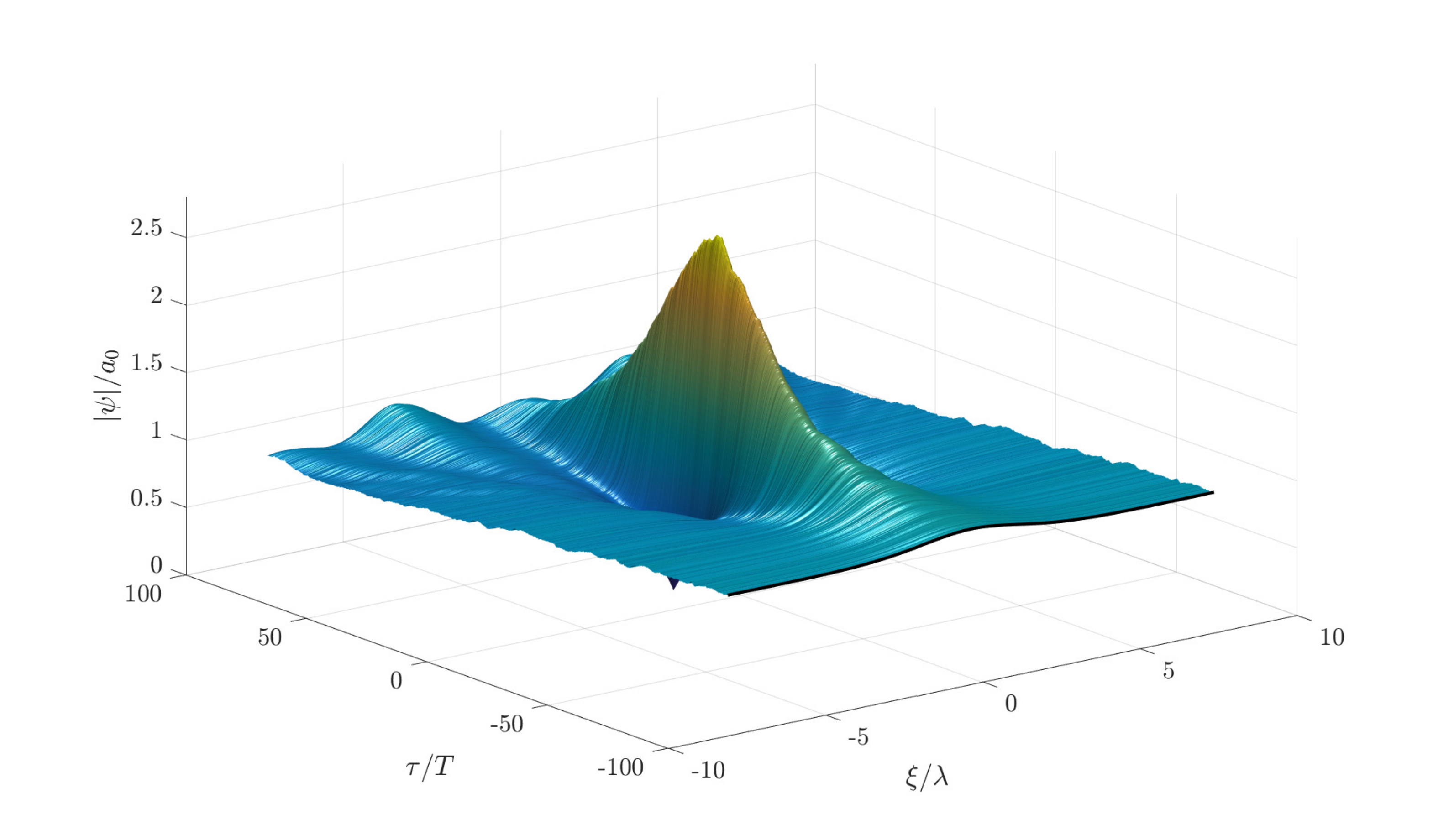}%
	\caption{Temporal evolution of a Peregrine solution of the stochastic NLS with white noise excitation in time, strong random forcing $\zeta_{\tau}$ with a variance \textcolor{black}{${\sigma^2= 0.16 \, \mathrm{1/s^2}}$}.}%
	\label{fig:whitw noise_PeregrineSolution}%
\end{figure}

A random Peregrine breather solution of the stochastic NLS~\eqref{eq:nlsnoise} is obtained, if initial conditions are used, which lead to the Peregrine breather \eqref{eq:unforced breatherOnorato} in the undisturbed case.
For strong random time dependent forcing by white noise $\zeta(\tau)$ with variance \textcolor{black}{${\sigma^2= 0.16 \, \mathrm{1/s^2}}$}, the temporal evolution of the random Peregrine breather is shown in Figure~\ref{fig:whitw noise_PeregrineSolution}. Thereby the carrier wave amplitude and frequency are set to $a_0=1 \, \mathrm{m}$ and $\omega=0.8 \, \mathrm{rad/s}$, respectively. Although the random forcing is very severe, which means that it is of more than one order of magnitude stronger than typical random forcing due to extreme wind conditions, a Peregrine type solution is clearly identifiable, including a significant peak in the wave amplitude. This shows that the development of Peregrine breather solutions can not be prevented even by very extreme random forcing of the stochastic NLS~\eqref{eq:nlsnoise}.

\section{Random excitation by Wind}\label{sec:RandomWind}
Further research is necessary for the case of physical random wind excitation of surface gravity waves in order to obtain new insights and a better understanding of the formation of high and extreme waves in wind seas.
A major task in the presented study is the generation of a realistic random wind velocity process for nonlinear wind excited waves, which can be used in the NLS~\eqref{eq:nlsnoise}. 
The NLS for weakly nonlinear waves was excited by a deterministic model of the wind as given in Kharif et al. \cite{Kharif:2010} or Onorato and Proment~\cite{onorato2012approximate}.
%
%

The stochastic wind excitation leads to an NLS, which is parametrically excited by a stochastic process. 
This problem is analyzed for a logarithmic wind profile as given in equation \eqref{eq:logWindProfile}, \textcolor{black}{whereby the corresponding wind is unidirectional and} either random in space or random in time. To simplify the notation, only the generation of a time dependent process will be considered here. A space dependent process can be generated analogously. An important question is, whether the characteristic behavior of  special solutions of the NLS, such as the Peregrine breather or the soliton solution, survives in an environment with random forcing due to wind.

In the following the necessary theory for the generation of a stochastic process for the wind velocity is put together.
Van der Hoven has shown in \cite{van1957power}, that typical wind velocity spectra in the surface boundary layer have a spectral gap. The wind speed fluctuations above this spectral gap contain components in the range from seconds to minutes and even faster fluctuations, which represent wind gusts. Such wind velocity processes can be described by the von Karman model, which is characterized by the power spectral density \cite{leithead1991role} 
\begin{equation}\label{eq:KarmanSpectrum}
S(\omega)=\frac{K_v}{(1+\omega^2T_v^2)^{5/6}}
\end{equation}
%
For the von Karman spectrum the coefficients $K_v$ and $T_v$ are given by
\begin{equation}\label{eq:K_v}
\begin{aligned}
K_v=0.475\, \sigma_v^2\, T_v,\\
T_v=\frac{L_v}{V_m}
\end{aligned}
\end{equation}
and depend on the mean wind speed $V_m$, the correlation length $L_v$, and the standard deviation $\sigma_v$ of the wind speed fluctuation.

A CARMA process as given in \textcolor{black}{Appendix~3} is used in order to generate a non white wind velocity process, which takes random wind gusts into account. 

\begin{figure}[htb]
	\centering	
	\includegraphics[width=0.9\columnwidth]{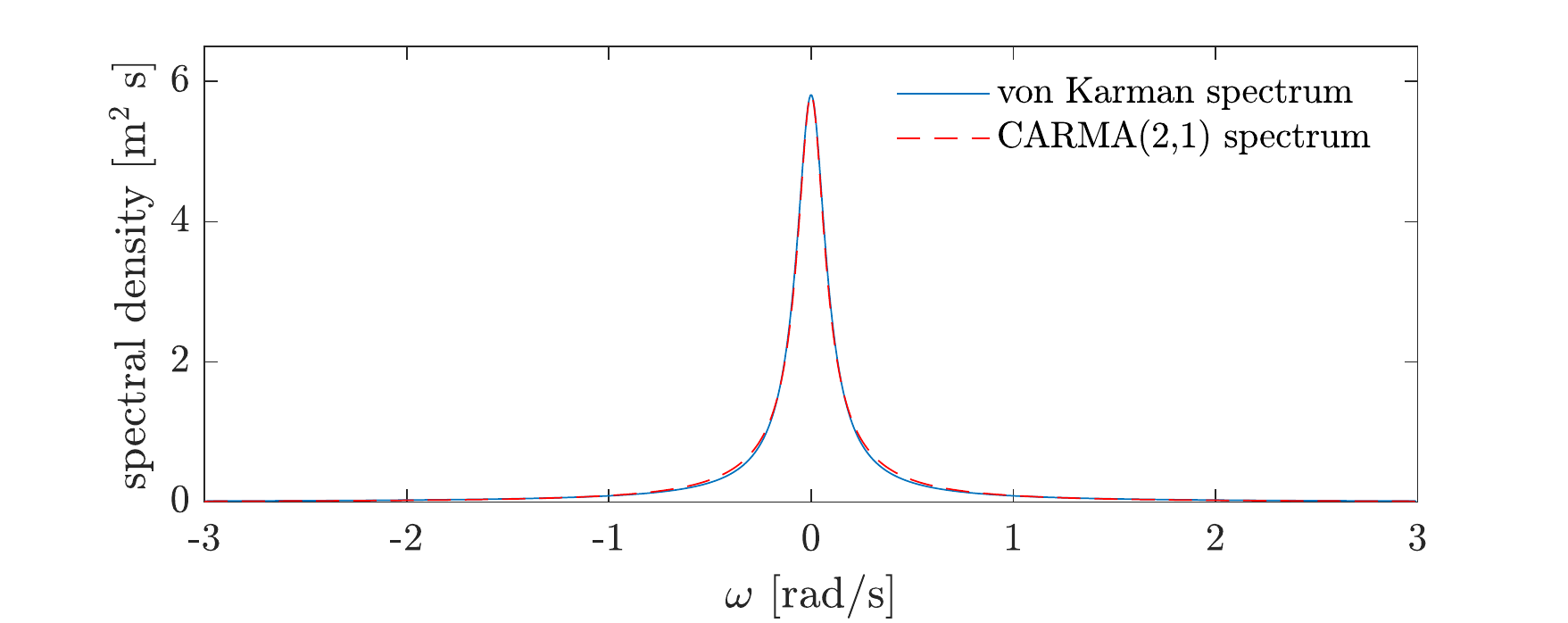}%
	\caption{Von Karman spectrum and its CARMA(2,1) approximation for mean wind speed ${V_m=50}$~km/h, correlation length  $L_v=170\, \mathrm{m/s}$, and $\sigma_v=1$.}%
	\label{fig:KarmanSpectrumCARMA21}%
\end{figure}

\begin{figure}[htb]
	\centering	
	\includegraphics[width=1.05\columnwidth]{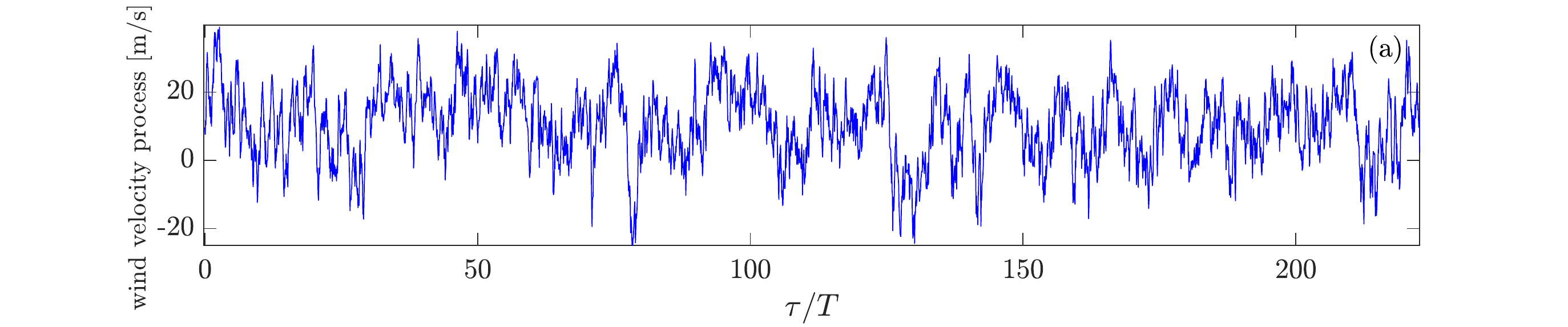}\\%
	\includegraphics[width=1.05\columnwidth]{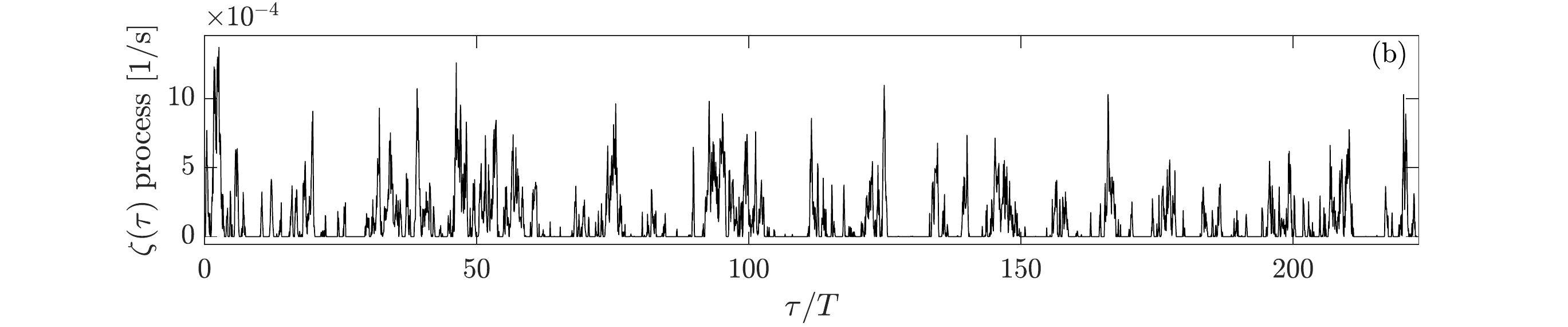}%
	\caption{ CARMA(2,1) wind velocity process ranging from -24 to 39 [m/s] \textcolor{black}{at constant height $z=50$ m} and corresponding time dependent random forcing process $\zeta(\tau)$.}%
	\label{fig:CARMA21WindProcess}
\end{figure}

A second-order rational transfer function approximation for the von Karman spectrum is chosen, which has been obtained in \cite{nichita2002large}
\begin{equation}\label{eq:LapCARMA21}
H_F(s)=K_v\frac{0.4\, T_v\,s+1}{(T_v\,s+1)(0.25s\,T_v+1)}.
\end{equation}
From this, a CARMA(2,1) process is generated, which is given by the following stochastic differential equation
\begin{equation}\label{eq:Carma21differentialgleichungssystem}
\begin{aligned}
y&=u_1,\\
\mathrm{d}u_1&=(u_2-a_1 u_1)\: \mathrm{d}\tau+b_1 \mathrm{d}W_{\tau},\\
\mathrm{d}u_2&=-a_2 u_1 \mathrm{d}\tau+b_0 \mathrm{d}W_{\tau},
\end{aligned}
\end{equation}
where $\mathrm{d}W_{\tau}$ is the increment of a standard Wiener process and 
\begin{equation}\label{eq:Carma21parameter}
\begin{aligned}
b_0=4 \sqrt{K_v}/T_v^2,\;\;\; b_1=1.6 \sqrt{K_v}/T_v,\;\;\; a_1=5/ T_v,\;\;\; a_2=4/T_v^2.
\end{aligned}
\end{equation}
%
%
%
%
%
%
%
%
%
%
This CARMA(2,1) process has the spectral density 
\begin{equation}\label{eq:ARMA21Spektraldichte}
S_F(\omega)=H_F(s)H_F(-s),
\end{equation}
which is an accurate approximation of the von Karman spectrum in equation \eqref{eq:KarmanSpectrum}, as can be seen in Fig.~\ref{fig:KarmanSpectrumCARMA21}.
A continuous time wind velocity process in the surface boundary layer can now be generated using the CARMA(2,1) process from equation \eqref{eq:Carma21differentialgleichungssystem}. An example of such a process is shown in Fig.~\ref{fig:CARMA21WindProcess}.

\section{Excitation of a soliton and a Peregrine breather by a random wind process}\label{sec:solitonPeregrineRandomWind}
In this section the stochastic NLS \eqref{eq:nlsnoise} is excited by a random wind velocity process $U(\xi,z,\tau)$ at height $z$, which is either time or space dependent, whereby $U(\xi,z,\tau):=u_1$ is generated by a CARMA(2,1) process from equation~\eqref{eq:Carma21differentialgleichungssystem}. A logarithmic wind profile $U(\xi,z,\tau)$ according to equation~\eqref{eq:logWindProfile} is assumed. From the randomly time or space varying wind velocity process $U(\xi,z,\tau)$ the friction velocity $u_*$ is calculated by means of a fix point iteration using equation~\eqref{eq:logWindProfile}. The resulting random friction velocity $u_*(\xi,\tau)$ defines the stochastic process $\zeta(\xi,\tau)$ resulting from equation~\eqref{eq:Gamma_neu}. Then this process is used as the excitation in the stochastic NLS~\eqref{eq:nlsnoise}. The necessary numerical steps are summarized in Procedure 1. \textcolor{black}{ In the following numerical calculations,  the coordinate system moving with the group velocity $c_g$ is considered.} \\

\begin{algorithm}[H]
	\nl generate turbulent wind velocity process from von Karman spectrum using equation \eqref{eq:Carma21differentialgleichungssystem} with coefficients from equation \eqref{eq:Carma21parameter}\\
	\nl determine $u_*$ from wind velocity process at height $z$ by fix point iteration using equation \eqref{eq:logWindProfile}\\
	\nl choose initial condition $\psi(\xi,0)=\psi_0(\xi)$ and calculate the stochastic partial differential equation \eqref{eq:nlsnoise} using the relaxation scheme as described in Appendix~\textcolor{black}{2}
	\caption{\textcolor{black}{Excitation of random sea waves}}\label{procedure}
\end{algorithm}



\subsection{Soliton under random time dependent wind excitation}
The effect of the turbulent wind on the soliton solution, which is a simple but fundamental solution of the NLS, can be seen in the following results.
A randomly time dependent wind forced soliton is calculated according to Procedure 1 using the undisturbed soliton solution from equation~\eqref{eq:solitonSolution} as initial condition and parameter values from  equations~\eqref{eq:Carma21parameter} and Table~\ref{tab:parameter}. 

\begin{figure}[htb]
		\centering	
		\includegraphics[width=1.05\columnwidth]{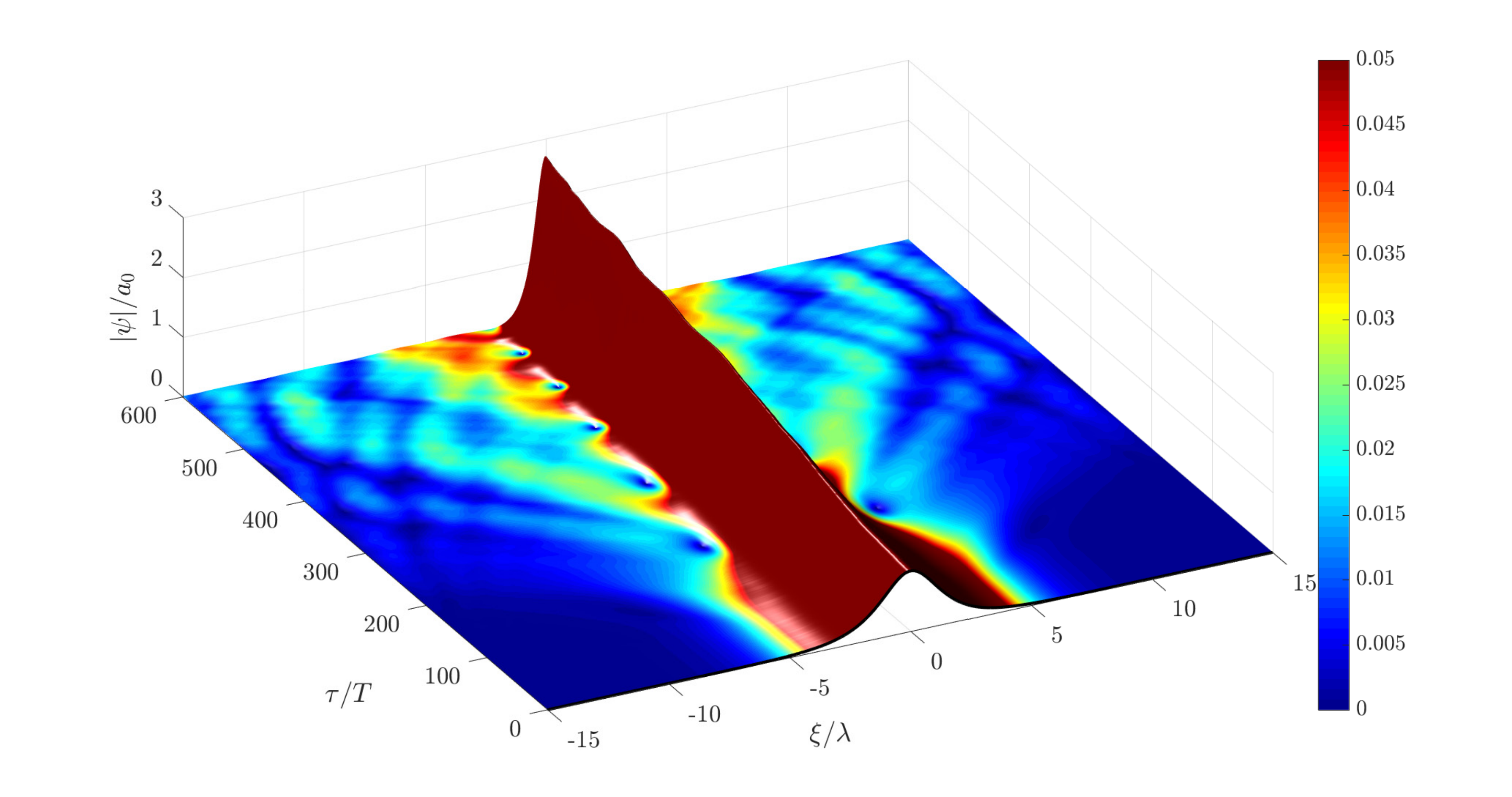}%
		\caption{Temporal evolution of a soliton solution of the stochastic NLS with non-white noise excitation in time, mean wind velocity $V_m=50$ km/h.}%
		\label{fig:nonwhite noise blowupsolitonSolution}%
	\end{figure}

A sample result for the case of a random wind excitation with mean wind velocity ${V_m= 50 \, \mathrm{km/h}}$ at the height $z=50\,\mathrm{m}$ is shown in Fig.~\ref{fig:nonwhite noise blowupsolitonSolution}. 
As can be observed in this figure, the zero water level gets disturbed as well during the evolution of the randomly wind forced soliton.
In contrast to the deterministic soliton solution in Fig.~\ref{fig:soliton}, a slightly time varying growth in the envelope amplitude and a symmetric \textcolor{black}{behavior} of the resulting random soliton solution in space $\xi$ is observed in Fig.~\ref{fig:nonwhite noise blowupsolitonSolution}. 
An important observation from the result in Fig.~\ref{fig:nonwhite noise blowupsolitonSolution} is, that random fluctuations in the wind excitation do not destroy the solitonic structure of the solution.

%

\subsection{Peregrine breather under random time dependent wind excitation}
Another important problem is the behavior of breather solutions if excitation by a random time dependent wind velocity process is applied.
%
%
Following Procedure 1, a numerical solution of the NLS \eqref{eq:nlsnoise} under random excitation by wind is obtained, using the parameter values from equation~\eqref{eq:Carma21parameter} and Table~\ref{tab:parameter}. Thereby, the carrier wave parameters are $a_0=1$ m and $\omega=0.8$ \textcolor{black}{rad/s} and the initial condition $\psi_0(\xi)$ is set according to equation~\eqref{eq:unforced breatherOnorato}, such that a Peregrine solution would develop in the deterministic case.

%
%



%


\begin{table}[!ht]
	\begin{center}
		{\small \caption{Parameter specifications.}\label{tab:parameter}\vspace{2mm}
			\begin{tabular*}{0.985\textwidth}{|c|c|c|c|}\label{tab:PendulumParameter}
				von K\'{a}rm\'{a}n constant $\kappa$  & air density $\rho_a$  &  water density $\rho$ & fluid viscosity $\nu$    \\
				\hline
				$0.4$    &1.225 [kg/ $\mathrm{m^3}$]&  1026.0 [kg/ $\mathrm{m^3}$] &$10^{-1}$ [$\mathrm{cm^2/ s}$] \\
			\end{tabular*}}
		\end{center}
	\end{table}
	
	%
	%
	%
	%
	%
	%
	%
	%
	%

	%
	%
	%
	%

\begin{figure}[htb]
		\centering	
		\includegraphics[width=1.0\columnwidth]{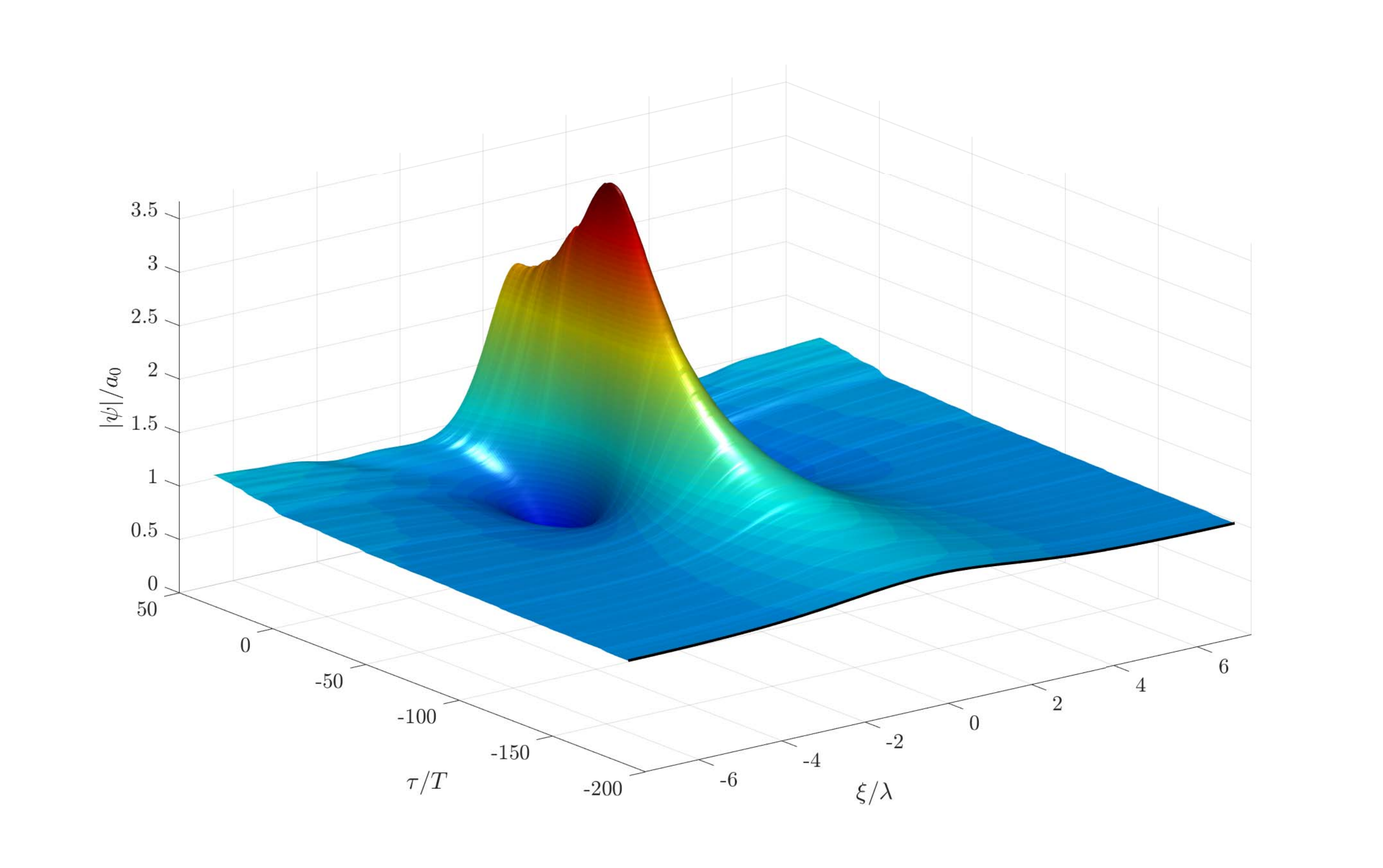}%
		\caption{Temporal evolution of a Peregrine solution of the stochastic NLS with non-white noise excitation in time for the random wind excitation from Fig.~\ref{fig:CARMA21WindProcess}.}%
		\label{fig:vonKarmanV50_PeregrineSolution}%
	\end{figure}

	Then, a sample solution of the NLS under strong random wind conditions with mean wind velocity $V_m = 50\, \mathrm{km/h}$ is obtained and a Peregrine breather can still be clearly identified, as shown in Fig.~\ref{fig:vonKarmanV50_PeregrineSolution}. 
	This solution has a peak amplitude, which is about three times larger than the mean amplitude of the other waves, indicating an extreme or freak wave.
	The results show, that a realistic excitation by random time dependent wind loads does not prevent breather solutions to occur even in strong wind conditions.
	
	
\begin{figure}[htb]
		\centering	
		\includegraphics[width=1.05\columnwidth]{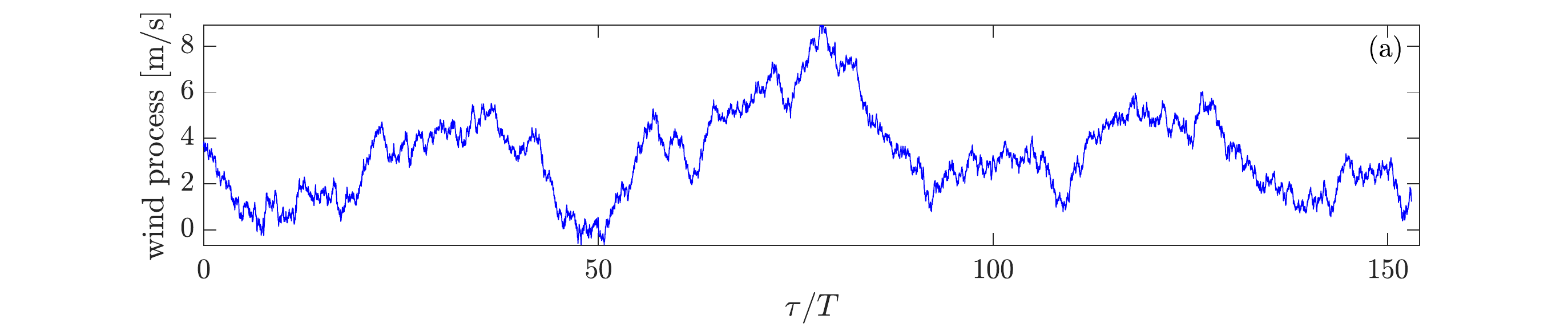}\\%
		\includegraphics[width=1.05\columnwidth]{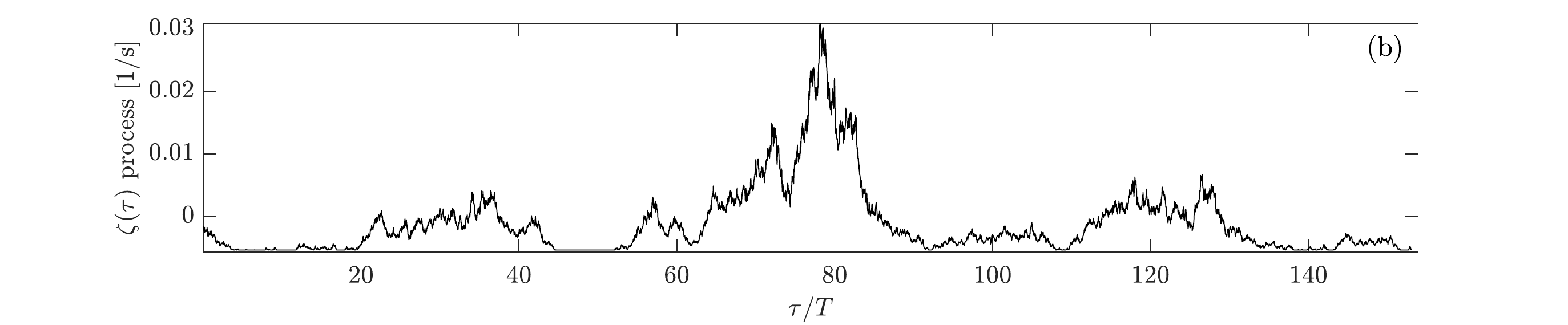}%
		\caption{ CARMA(2,1) wind velocity process at $z=1$ m with $V_m=5$ m/s, ranging from 0 to 8 [m/s], and corresponding random time dependent forcing process $\zeta(\tau)$.}%
		\label{fig:CARMA21WindProcess_Experiment}
	\end{figure}

\begin{figure}[htb]
		\centering	
		\includegraphics[width=1.0\columnwidth]{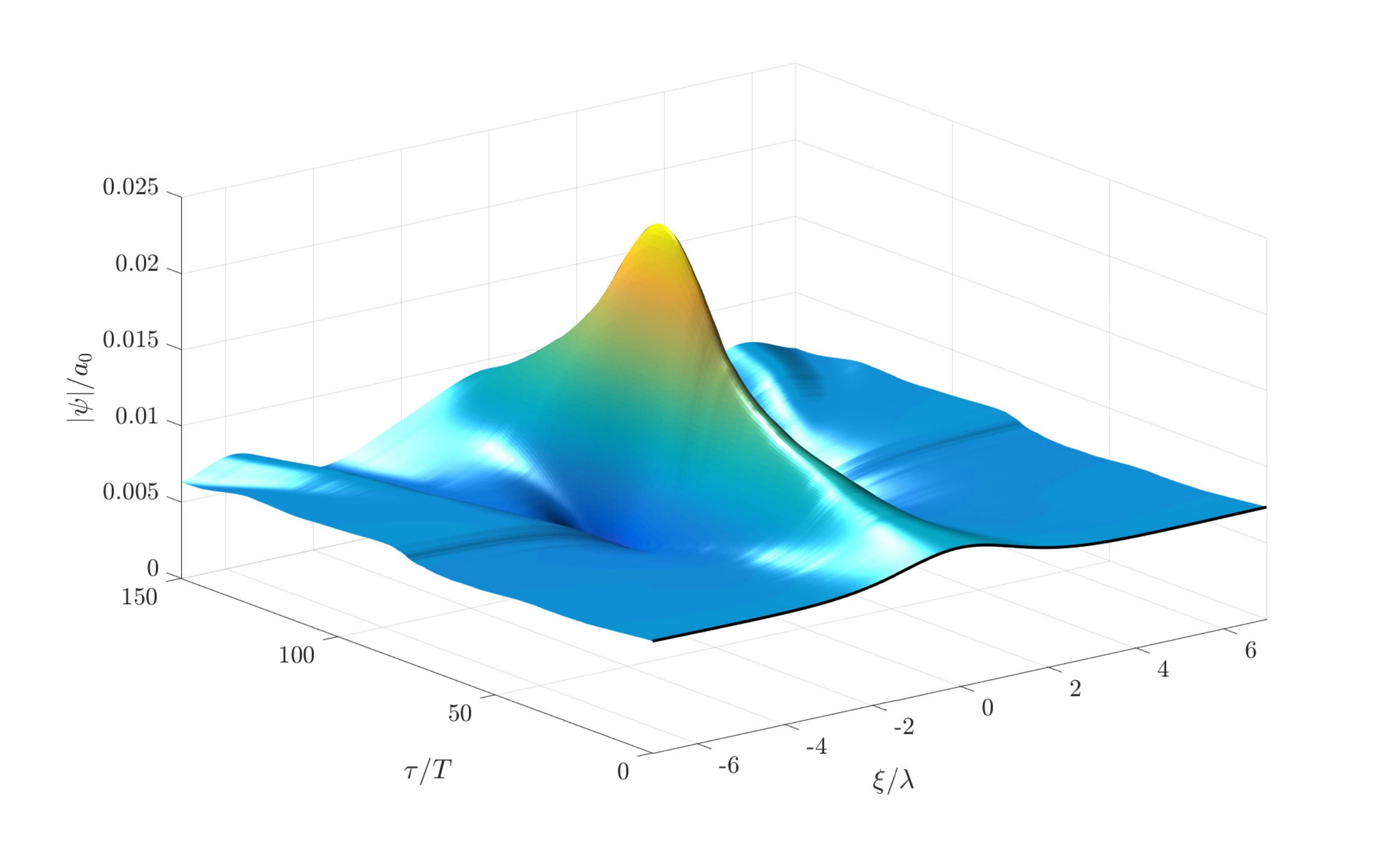}%
		\caption{Temporal evolution of a laboratory scale Peregrine solution of the stochastic NLS for the random wind excitation from Fig.~\ref{fig:CARMA21WindProcess_Experiment} with $V_m=5$ m/s and the carrier parameters $a_0=0.75$ cm, $\omega=10.68$ rad/s, $k=11.64$ \textcolor{black}{1/m}.}%
		\label{fig:vonKarmanV50_PeregrineSolutionExperiment}%
	\end{figure}

	The experimental studies presented in \cite{chabchoub2013experiments} on the effect of wind forcing on the modulation instability and the Peregrine breather forced by wind with velocities from 3 m/s to 9 m/s deal with a carrier wave  amplitude of ${a_0=0.75 \, \mathrm{cm}}$ and frequency $\omega=10.68$ \textcolor{black}{rad/s}.
	For the same carrier wave parameters and a random wind forcing ranging from 0 to 8 m/s, as shown in Fig.~\ref{fig:CARMA21WindProcess_Experiment}, a Peregrine breather type solution of the stochastic NLS is calculated and shown in Fig.~\ref{fig:vonKarmanV50_PeregrineSolutionExperiment} .
	
	By comparing the random processes $\zeta(\tau)$ in Figures~\ref{fig:CARMA21WindProcess} and \ref{fig:CARMA21WindProcess_Experiment}, it can be seen that the viscosity $\nu$ has a much greater influence at the experimental scale, since there is a significantly greater amount of time at which the process $\zeta(\tau)$ is negative. Moreover, in comparison to the case in Figure~\ref{fig:vonKarmanV50_PeregrineSolution}, the random forcing process is much stronger. Hence, the resulting forced Peregrine solution is much more perturbed but still clearly identifiable.
	
	It is stated in the experimental study \cite{chabchoub2013experiments} that the Peregrine breather has been also detected under strong wind conditions and that the observations are in line with the analytical results in equation~\eqref{eq:forced breatherOnorato} as derived in \cite{onorato2012approximate}. Furthermore, these experimental results validate that the effect of strong winds on the modulation instability, and hence on the development of breather solutions, is small.
	This is also confirmed by our results, where a Peregrine type solution for wind velocities from 0 to 8 m/s can always be identified.
	Considering the above results\textcolor{black}{, we can conclude} that Peregrine breathers can still be seen as prototypes of rough waves in the presence of random wind forcing.
	
	In the presented study, weakly nonlinear water waves forced by random wind are analyzed only in one spatial dimension. However, random changes in the wind direction are important as well. Therefore, future research on random perturbation of nonlinear water waves by wind should include the evolution equations of the wave envelope in two spatial dimensions, which leads to the analysis of randomly perturbed Davey-Stewardson equations for the case of arbitrary water depth.

	\subsection{Peregrine breather under random space dependent wind excitation}   
\begin{figure}[htb]
		\centering	
		\includegraphics[width=1.05\columnwidth]{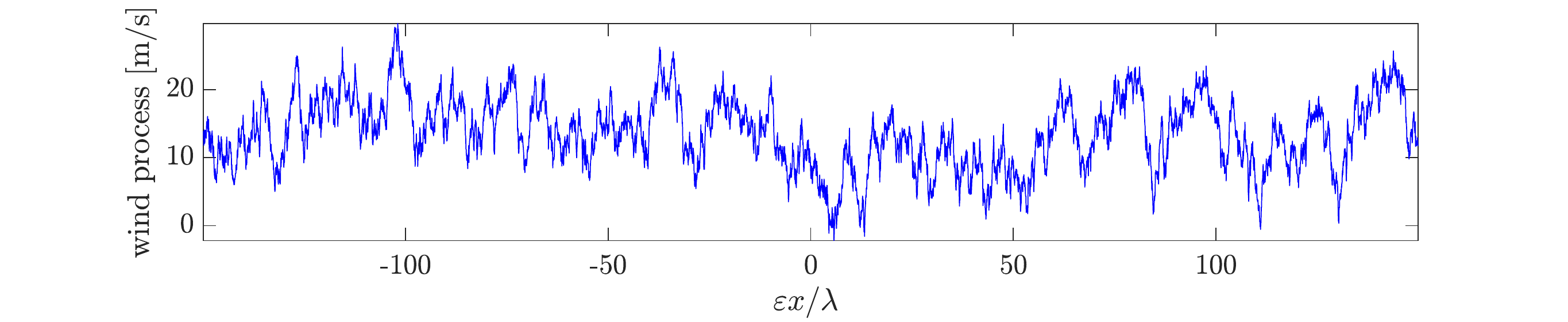}\\%
		\includegraphics[width=1.05\columnwidth]{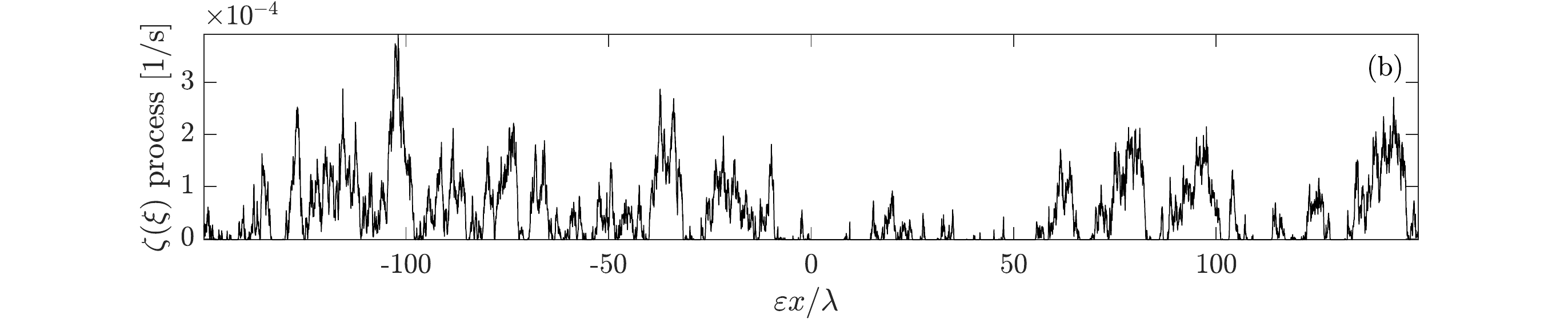}%
		\caption{ CARMA(2,1) wind velocity process in space and corresponding random space dependent forcing process $\zeta(\xi)$ with $V_m=50\, \mathrm{km/h}$ \textcolor{black}{at constant height $z=50$ m}}%
		\label{fig:CARMA21WindProcess_space}
	\end{figure}

	So far a time dependent wind forcing is considered in the preceding subsections. In addition to these results, a space dependent excitation and its influence on the behavior of solutions of \eqref{eq:nlsnoise} is investigated. 
	Using the CARMA(2,1) process from equation ~\eqref{eq:Carma21differentialgleichungssystem}, the parameter values of equation ~\eqref{eq:Carma21parameter}, Table~\ref{tab:parameter} and the mean wind velocity $V_m=50\, \mathrm{km/h}$, a stochastic process as shown in Fig.~\ref{fig:CARMA21WindProcess_space} can be generated. Again, the carrier wave parameters are $a_0=1$ m, ${\omega=0.8\, \mathrm{rad/s}}$, and the initial condition $\psi_0(\xi)$ is set according to equation~\eqref{eq:unforced breatherOnorato}, such that again a Peregrine solution would develop in the deterministic case. 
	
	\textcolor{black}{At this point, it has to be mentioned that the considered numerical solutions depend on the coordinate-system ($\xi,\tau$), which moves in contrast to the space-time coordinate system ($x$,$t$) with the group velocity $c_g$. In order to compute a numerical solution under a space dependent wind forcing, in each timestep the space dependent wind is generated inside a frame which moves with the group velocity $c_g$. This means that a wind is considered, which is variable in time and space, whereby time and space are linear dependent.}
	
\begin{figure}[htb]
		\centering	
		\includegraphics[width=1.05\columnwidth]{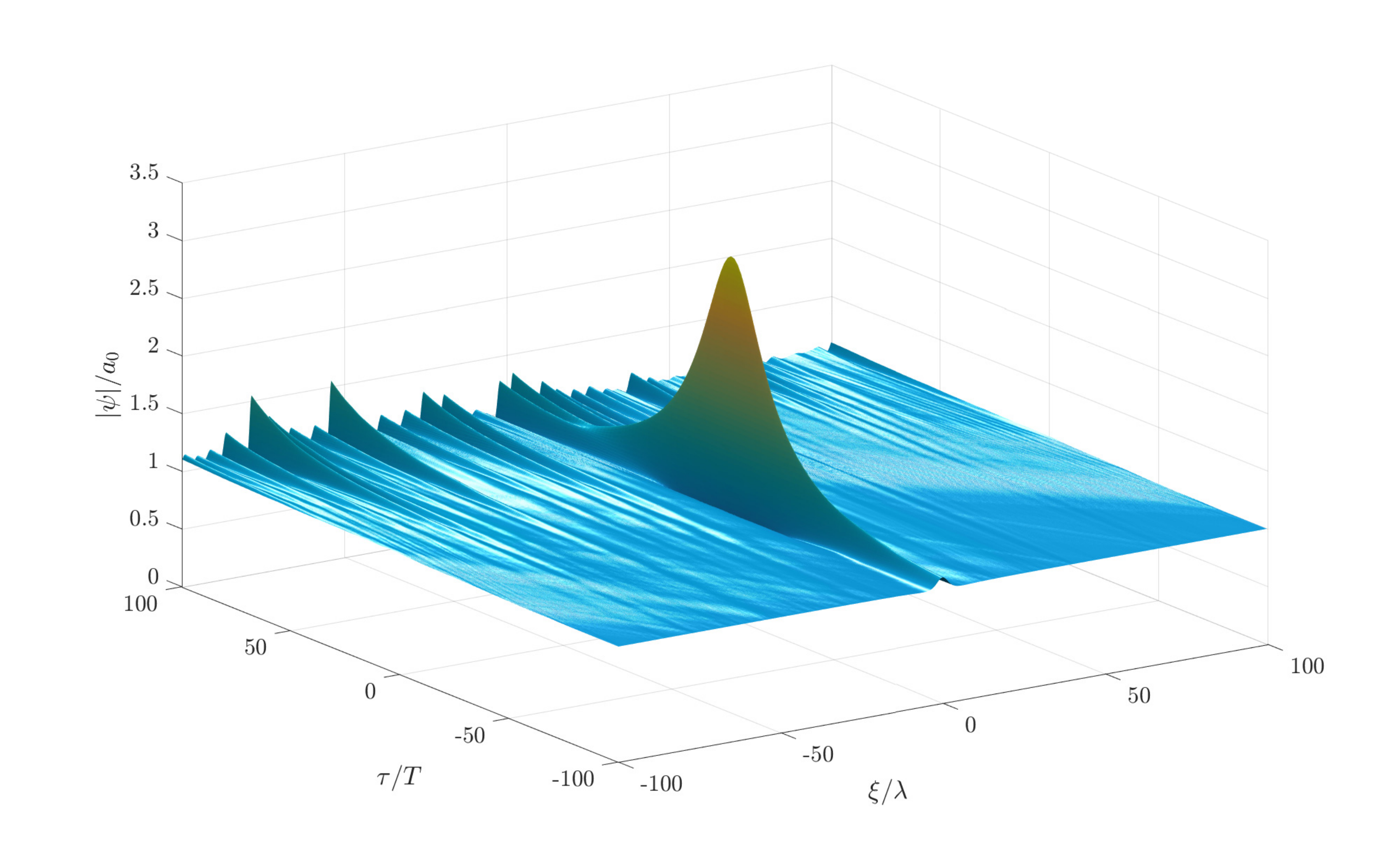}%
		\caption{Evolution of a Peregrine solution of the stochastic NLS with a non-white noise excitation in space, mean wind velocity $V_m=50\, \mathrm{km/h}$.}%
		\label{fig:Peregrine_space_fullsimulation}%
	\end{figure}

	Following Procedure 1, the numerical solution shown in Fig.~\ref{fig:Peregrine_space_fullsimulation} is obtained. As can be seen, the general structure of the Peregrine solution remains for the considered excitation. However in contrast to the case of time dependent random excitation, fluctuations in space appear, which make the solution asymmetric. Nevertheless, it can again be concluded that Peregrine breathers can be identified in the case of space dependent random wind forcing and on the basis of these results are not excluded as prototypes of rough waves. 
	
	\section{Conclusion}
	Until now there is doubt whether prototypes of extreme waves, like the Peregrine breather, exist also in open seas with randomly distributed wave heights and forcing due to turbulent wind.
	There are results in \cite{antoine2015gpelab,antoine2015modeling} on nonlinear Schr\"odinger equations with a potential, which is disturbed by white noise.
	Moreover, de Bouard et al. \cite{Bouard1999,DeBouard2001} studied a version of the NLS, which is disturbed by an additive and multiplicative real white noise. However, for the randomly wind excited NLS, the excitation is non-white, purely imaginary, and multiplicative. 
	This case is treated in the present work and the resulting impact on soliton and Peregrine breather solutions is analyzed. 
	As a result it is found that breathers can exist in deep water even in the presence of random wind excitation. 
	In addition, strong random excitation of the NLS by white noise is considered. In this case the Peregrine breather is clearly identifiable as well. Thus also in such extreme cases, the development of breathers is not prevented.
	Therefore, the obtained results indicate that Peregrine breather and soliton solutions can exist in a randomly forced environment like the oceans. 

 \section*{Appendix 1\,\,\, Derivation \,of\, \,the \,\,forced \, nonlinear\\ 
 	Schr\"odinger equation for time and space variant wind-induced pressure}\label{Appendix:Multiple_scales}
 It can be shown that weakly nonlinear solutions of the Euler equations can be reduced to solutions described by a complex envelope, which satisfies the NLS. Such reduction can be achieved by the method of multiple scales, cf.~\cite{Mei:1983}. 
 In the following the multiple scales analysis is presented for the case of forcing of water waves by a time and space variant wind field.
 For the forced Euler equations, an incompressible fluid with density $\rho$, a finite depth $h$, and a free surface $\eta(x,t)$ is assumed, where $x$ is the spatial variable and $t$ is time, and the corresponding velocity field is irrotational. This leads to four equations, 
 where the nonlinear gravity wave problem is governed by two linear equations, which are the Laplace equation and the kinematic condition at the sea bed, and by kinematic and dynamic free surface conditions, which are nonlinear~\cite{osborne:2010}. 
 If in addition a time and space variant wind-induced pressure $P_a(x,t)$ at the free surface is assumed, and if dissipative effects are included by means of the water kinematic viscosity $\nu$, \textcolor{black}{as suggested} by Dias~et~al.~\cite{dias2008}, then the forced Euler equations become 
 \begin{subequations}\label{eq:Euler_equations}
 	\begin{align}
 	\phi_{xx}+\phi_{zz}&=0,    \hspace{0mm}&\mathrm{for}&\hspace{8mm}-h\leq z\leq \eta(x,t),\label{eq:ns1}~\\
 	\eta_t+\phi_x\eta_x-\phi_z-2\nu\eta_{xx}&=0,\hspace{0mm}  &\mathrm{for}&\hspace{8mm}z=\eta(x,t),\label{eq:ns3}~\\
 	\phi_t+\frac{1}{2}\left( \phi_x^2+\phi_z^2\right)+g\,\eta&=-\frac{1}{\rho}P_a-2\nu \phi_{zz},\hspace{0mm}  &\mathrm{for}&\hspace{8mm}z=\eta(x,t),\label{eq:ns4}~\\
 	\phi_{z}&=0,  \hspace{0mm} &\mathrm{for}&\hspace{8mm}z=-h,\label{eq:ns2}~
 	\end{align}
 \end{subequations}
 where  $\phi(x,z,t)$ is the velocity potential and $g$ is acceleration due to gravity.


 The nonlinear boundary conditions at the free surface $\eta(x,t)$ make the solution of equations~\eqref{eq:Euler_equations} very difficult, since the nonlinear boundary $\eta(x,t)$ is unknown. Thus, one seeks to use simplifications.

 
 Here, the method of multiple scales  is used for the derivation of a nonlinear Schr\"odinger equation resulting from the forced Euler equations \eqref{eq:Euler_equations}. 
 Following Dawey and Stewardson \cite{davey1974three} and Hasimoto and Ono \cite{hasimoto1972nonlinear}, the scalings 
 \begin{equation}
 \xi=\varepsilon (x-c_g\,t),\quad\quad \tau=\varepsilon^2\,t
 \end{equation}
 are used, whereby \textcolor{black}{the small parameter $\varepsilon$ represents the wave steepness,}
 $c_g=\dfrac{g}{2\omega}(\tanh kh+kh(1-\tanh^2kh))$ is the group velocity, $\omega$ is the frequency, and $k$ is the wave number of the considered carrier wave. 
 The velocity potential $\phi$ and the surface elevation $\eta$ are expanded in series of the form
 \begin{align} 
 &\phi(x,z,t)=\sum_{n=1}^\infty\varepsilon^n\sum_{m=-n}^n\phi^{n,m}(\xi,z,\tau)\,E^m,\label{eq:phi_ansatz}\\
 &\eta(x,t)=\sum_{n=1}^\infty\varepsilon^n\sum_{m=-n}^n\eta^{n,m}(\xi,\tau)\,E^m,\label{eq:eta_ansatz}
 \end{align}
 where  
 \begin{equation}
 E=\exp\left( \mathrm{i}(kx-\omega t) \right),\quad\quad \phi^{(n,-m)}=\bar{\phi}^{(n,m)},\quad\quad\eta^{(n,-m)}=\bar{\eta}^{(n,m)},\notag
 \end{equation}
 and a bar denotes the complex conjugate.
 As a next step, the velocity potential at the free surface $\eta$ is expanded in a Taylor series around $z=0$
 \begin{equation}\label{eq:TaylorExpansion}
 \phi(x,\eta,t)=\sum_{j=0}^\infty\frac{ \eta^j}{j!} \frac{\partial^j}{\partial z^j} \phi\bigg|_{z=0}
 \end{equation}
 Then the expansions \eqref{eq:phi_ansatz}-\eqref{eq:TaylorExpansion} are substituted into the equations \eqref{eq:Euler_equations}.
 As in \cite{leblanc2007amplification,Kharif:2010} the wind-induced pressure $P_a$ is assumed to be of order $\mathcal{O}(\varepsilon^3)$.
 Therefore, the wind-induced pressure evaluated at $z=0$ is expanded as
 \begin{align}
 P_a(x,t)=\sum_{n=1}^\infty\varepsilon^{n-1}\sum_{m=-n}^n p^{n,m}(\xi,\tau)\,E^m.\label{eq:p_ansatz}
 \end{align}
 and substituted into the boundary condition \eqref{eq:ns4}.
 Then, terms of linear and quadratic order are not affected by wind forcing and the well-known results for $\phi^{m,n}(\xi,z,\tau)$ with $n\leq2$ can be used \cite{davey1974three,hasimoto1972nonlinear}. These are given as\\
 \begin{equation}\label{eq:DSconditions}
 \begin{aligned}
 &\phi^{1,1}=\psi \frac{\cosh k(z+h)}{\cosh kh},\quad \quad \phi^{1,0}_z=0,\quad \quad \phi^{2,0}_z=0,\\ 	&\phi^{2,1}=D\frac{\cosh k(z+h)}{\cosh kh}-\mathrm{i} \beta_1  \psi_{\xi},	\\
 &\phi^{2,2}=\mathrm{i}\beta_2\frac{\cosh 2k(z+h)}{\cosh(2kh)},\quad  \quad \eta^{1,0}=0,	\quad  \quad g\eta^{1,1}= \mathrm{i} \omega \psi, \\
 &  g\eta^{2,0}=c_g \phi^{1,0}_{\xi}-k^2(1-\theta^2)|\psi|^2, \quad g\eta^{2,1}=\mathrm{i}\omega D+ c_g\psi_{\xi},	\quad  g\eta^{2,2}=\beta_3 \psi^2,		
 \end{aligned}
 \end{equation}	
 where 
 \begin{equation}
 \begin{aligned}
 &\beta_1= -\frac{(z+h)\sinh(k(z+h))-h\theta\cosh(k(z+h))}{\cosh(kh)},\\
 &\beta_2= \frac{3k^2(1-\theta^4(kh))}{4\omega\theta^2},\;	\beta_3=\frac{k^2(\theta^2-3)}{2\theta^2},\;	\theta=\tanh(kh),
 \end{aligned}
 \end{equation}	
 $\omega$ fulfills the dispersion relation $\omega=\sqrt{gk\theta}$, and \textcolor{black}{$\psi$ and $D$ are functions of the slow variables} $\xi$, $\tau$.
 
 Terms of the velocity potential $\phi$ of order $n=3$ and harmonic $m=0$ in boundary condition \eqref{eq:ns3} do not involve the wind-induced pressure $P_a$ or the viscosity $\nu$, and lead similarly as in \cite{davey1974three} to\\
 \begin{equation}
 \phi_{\xi}^{1,0}=\beta_4|\psi|^2,\; \mathrm{where}\quad \beta_4=-k^2\frac{ 2c_p+c_g(1-\theta^2) }{gh-c_g^2},\quad 	\phi_{\xi z}^{1,0}=0,
 \end{equation}
 and $c_p=\omega/k$ is the phase velocity.
 From the Laplace equation \eqref{eq:ns1} together with the bottom boundary condition \eqref{eq:ns2}, ${\phi}^{3,1}$ is obtained as
 \begin{equation}
 \begin{aligned}
 {\phi}^{3,1} = &\;\; \frac{\cosh k(z+h)}{\cosh kh}G+\frac{\beta_1}{2k}\left(2kh\theta^2\psi_{\xi\xi}-2\mathrm{i}k D_{\xi} \right)\\
 &-\frac{\left((z+h)^2-h^2\right)\cosh(k(z+h)}{2 \cosh(kh)}\psi_{\xi\xi},
 \end{aligned}
 \end{equation}
 whereby $G$ is a function of $\xi$, $\tau$.
 This is the same result as in \cite{djordjevic:1977} (without surface tension), since no wind-induced pressure is involved in the Laplace equation and the bottom boundary condition.
 Collecting terms of the velocity potential $\phi$ of order $n=3$ and harmonic $m=1$ in boundary condition~\eqref{eq:ns3}, the equation 
 \begin{equation}\label{eq:31bndaryCond1}
 \begin{aligned}
 \phi_{z}^{3,1}+\mathrm{i}\omega{\eta}^{3,1}=&-c_{g}\,\eta _{\xi }^{2,1}+\eta _{\tau}^{1,1}+\mathrm{i}k{
 	\eta}^{1,1}\phi _{\xi }^{1,0}+2\,\nu\,k^2\,{\eta}^{1,1}-{\eta}^{2,0}\phi_{\it zz}^{1,1}-\bar{\eta}^{1,1}\phi_{zz}^{2,2}\\
 &-\eta^{2,2}\bar{\phi}_{zz}^{1,1}+\left(\eta^{1,1}\right)^2 \left(k^2 \bar{\phi}_{zz}^{1,1}-\frac{1}{2} \bar{\phi}_{zzz}^{1,1}\right)
 -\eta^{1,1}\bar{\eta}^{1,1} \phi_{zzz}^{1,1}	\\
 &+2 k^2	\bar{\eta}^{1,1}\phi_{z}^{2,2}+2 k^2 {\eta}^{2,2}\bar{\phi}_{z}^{1,1}	
 \end{aligned}
 \end{equation}		
 is obtained. Thereby, terms that are zero in equations \eqref{eq:DSconditions} have already been omitted.
 The corresponding equation for $n=3$ and $m=1$ in boundary condition \eqref{eq:ns4} is given by
 \begin{equation}\label{eq:31bndaryCond2}
 \begin{aligned}
 -\mathrm{i}{\phi}^{3,1}\omega+g{\eta}^{3,1}=&\,-{\frac {1}{\rho_{w}}}{p}^{1,1}-2\,\nu\,{\phi}^{1,1}_{zz}-\frac{1}{2}\mathrm{i}\omega\,{\eta}^{2,2}{\bar{\phi}_{{{\it zz
 			}}}}^{1,1}-c_{g}\,{\phi}^{2,1}_{\xi }-{\phi}^{
 			1,1}_{\tau }\\
 		&-\mathrm{i}k{\phi}^{1,1}\phi _{\xi }^{1,0}-\mathrm{i}\omega\,{\eta}^{2,2}{
 			\bar{\phi}_{{z}}}^{1,1}-2\,{k}^{2}\bar{\phi}^{1,1}{
 			\phi}^{2,2}-{k}^{2}{\eta}^{1,1}{\phi_{{z}}}^{1,1}{\bar{\phi}}^{1,1}\\
 		&-{k}^{2}{\eta}^{1,1}{\bar{\phi}_{{z}}}^{1,1}{\phi}^{1,1
 		}+{k}^{2}{\bar{\eta}}^{1,1}\phi_{{z}}^{1,1}{\phi}^{1,1}+2\,\mathrm{i}
 		\omega\,{\bar{\eta}}^{1,1}\phi_{{z}}^{2,2}\\
 		&+\mathrm{i}\omega\,{\eta_{}}^{
 			1,1}{\bar{\eta}}^{1,1}\phi_{{{\it zz}}}^{1,1}-\phi_{{z}}^{
 			2,2}{\bar{\phi}_{{z}}}^{1,1}+\mathrm{i}\omega\,{\eta}^{2,0}\phi_{z}^{
 			1,1}\\
 		&-{\eta}^{1,1}\phi_{{z}}^{1,1}{\bar{\phi}_{{{\it zz}}}}^{1,1}-{\eta}^{1,1}{\bar{\phi}_{{z}}}^{1,1}\phi_{{{\it zz}}}^{1,1
 		}-{\bar{\eta}}^{1,1}\phi_{{z}}^{1,1}\phi_{{{\it zz}}}^{1,1}
 		,
 		\end{aligned}
 		\end{equation}		
 		and includes the leading order time and space varying wind-induced pressure term $p^{1,1}(\xi,\tau)$.		
 		It is also possible to assume $P_a$ to be of order $\mathcal{O}(\varepsilon^2)$, which involves additional terms due to the wind-induced pressure, as can be seen in \cite{brunetti2014nonlinear}, where a time and space invariant wind velocity is considered.
 		\textcolor{black}{Subtracting \eqref{eq:31bndaryCond1} from \eqref{eq:31bndaryCond2} such that the unknown ${\eta}^{3,1}$ vanishes}
 		leads finally to the evolution equation for the wave envelope 
 		$\psi(\xi,\tau)$, which is subjected to a 
 		time and space variant wind-induced pressure. This equation is given by 
 		\begin{equation}\label{eq:generalNLS}
 		\mathrm{i}\,\psi_{\tau}+\mu\,\psi_{\xi\xi}-\gamma\,|\psi|^2\psi=-\mathrm{i}\,\frac {p^{1,1}}{2\rho_{w}}-\mathrm{i}\,2\,\nu\,k^2\,\psi,
 		\end{equation}
 		where $\mu=\frac{1}{2}\frac{\partial^2}{\partial k^2} \omega(k)$, and
 		\begin{equation}
 		\begin{aligned}
 		\gamma=&\frac{k^4}{4\omega}\left(\frac{9}{\theta^2}-12+13\theta^2-2\theta^4-\frac{2\left( 4c_p^2+4c_p c_g(1-\theta^2)+c_g^2(1-\theta^2)^2 \right)}{gh-c_g^2}  \right).\notag
 		\end{aligned}
 		\end{equation}
 		The dissipative term $-\mathrm{i}\,2\,\nu\,k^2\,\psi$ in equation \eqref{eq:generalNLS} is the same as was for example previously obtained in \cite{carter2016frequency}.           
 		It is noted, that only the space and time variant pressure component $p^{1,1}(\xi,\tau)$ from the pressure expansion \eqref{eq:p_ansatz} is involved in this equation.
 		For deep water, \eqref{eq:generalNLS} reduces \textcolor{black}{with $\theta \rightarrow 1$, $c_g \rightarrow \frac{\omega}{2k}$ and a scaling of $\psi \rightarrow \frac{\omega}{2k} \psi$} to
 		\begin{equation}\label{eq:generalNLS2}
 		\mathrm{i}\,\psi_{\tau}-\frac{\omega}{8 k^2}\,\psi_{\xi\xi}-\frac{1}{2}\omega k^2\,|\psi|^2\psi=-\mathrm{i}\,k\,\frac { p^{1,1}}{\omega\rho_{w}}-\mathrm{i}\,2\,\nu\,k^2\,\psi.
 		\end{equation}

 		 \section*{Appendix 2 Relaxation method for nonlinear gravity waves}\label{sec:numericalMethods}
 		%
 		%
 		%
 		%
 		Using the relaxation scheme, the deterministic NLS \eqref{eq:nlsdeterministic} is discretized by \cite{antoine2015gpelab,antoine2015modeling}
 		\begin{equation}\label{eq:relaxation_deterministic}
 		\begin{aligned}
 		&\frac{\phi^{n+1/2}+\phi^{n-1/2}}{2}=\frac{1}{2}\omega\,k^2|\psi^n|^2,\\
 		&\mathrm{i}\frac{\psi^{n+1}-\psi^n}{\delta t}=\frac{\omega}{8k^2} \, \left(\frac{\psi^{n+1}+\psi^n}{2}\right)_{xx}+\phi^{n+1/2}\frac{\psi^{n+1}+\psi^n}{2}\\
 		&\hspace{2.2cm}+\frac{\Gamma^{n+1}\,\psi^{n+1}+\Gamma^n\,\psi^n}{2}.
 		\end{aligned}
 		\end{equation}
 		Moreover, the space can be discretized by a pseudo-spectral approximation, which leads to a high accuracy, see for example \cite{antoine2015modeling}.
 		For the white noise case, where the excitation $\zeta$ of the stochastic NLS \eqref{eq:nlsnoise} is given by a scalar time varying white noise $\zeta=\zeta(t)$ as specified in Definition~\ref{def:CARMApqZustandsform}, the stochastic integral is approximated in the sense of Stratonovich by
 		\begin{equation}
 		\int_{t_n}^{t_{n+1}}\psi(s,x)\circ\mathrm{d}W(s)  \approx \frac{\psi(t_{n+1},x)+\psi(t_{n},x)}{2}\, ( W_{t_{n+1}}- W_{t_n}),
 		\end{equation}
 		where $\mathrm{d}W(t)=\zeta(t)\, \mathrm{d}t$ is the increment of the standard Wiener process $W_t$. 
 		
 		We set {${\chi^{n+1/2}:=( W_{t_{n+1}}- W_{t_n})/\sqrt{\delta t}}$}, which are distributed according to the normal distribution $\mathcal{N}(0,1)$.
 		Then the relaxation scheme for the NLS, which is excited by white noise in time, reads
 		\begin{equation}\label{eq:relaxation_white}
 		\begin{aligned}
 		&\frac{\phi^{n+1/2}+\phi^{n-1/2}}{2}=\frac{1}{2}\omega\,k^2|\psi^n|^2,\\
 		&\mathrm{i}\frac{\psi^{n+1}-\psi^n}{\delta t}=\frac{\omega}{8k^2} \, \left(\frac{\psi^{n+1}+\psi^n}{2}\right)_{xx}+\phi^{n+1/2}\frac{\psi^{n+1}+\psi^n}{2}\\
 		&\hspace{2.2cm}+\frac{1}{2\sqrt{\delta t}}\chi^{n+1/2}(\psi^{n+1}+\psi^n).
 		\end{aligned}
 		\end{equation}
 		For the above numerical calculations, the GPELab toolbox can be used \cite{antoine2015gpelab}.
 		In the case, where the NLS is excited by a random wind process in time, we first have to generate the non-white wind velocity process $U(z,t_n)$ by means of a CARMA process, as given in Definition \ref{def:CARMApqZustandsform}. This can be done using for example the Euler-Maruyama scheme \cite{kloeden:1992}. From the randomly time varying wind velocity process $U(z,t_n)$ the friction velocity $u_*(t_n)$ is calculated by means of a fix point iteration at each instant of time using equation~\eqref{eq:logWindProfile}. The resulting random friction velocity $u_*(t)$ defines the stochastic process $\Gamma^n$, which is then substituted into equation \eqref{eq:relaxation_deterministic} where
 		\begin{equation}
 		\Gamma^n:=\frac{k\omega}{2g}\frac {\rho_a}{\rho_{w}} \,\beta\,\left(\frac{u_*(t_n)}{\kappa}\right)^2  -2\,\nu\,k^2.
 		\end{equation}

 		 \section*{Appendix 3 CARMA process}\label{Appendix:CARMA}
 		A CARMA process is defined as follows, cf.~\cite{brockwell:1995,dostal:2011}.
 		\begin{definition}\textbf{(CARMA(p,q) process)}\label{def:CARMApqZustandsform}\\
 			A CARMA(p,q) process $y(t)$, $0\leq q< p$, is defined as the stationary solution of
 			\begin{equation}\label{eq:DGLCARMApqZustandsform}
 			y=\mathbf{c}\;\mathbf{u}(t),
 			\end{equation}
 			with the linear differential equation for the state vector
 			$\mathbf{u}(t)\in \mathds{R}^p$
 			\begin{equation}\label{def:CARMApqZustandsformDGL}
 			\mathbf{\dot{u}}(t)=\mathrm{\mathbf{A}}\;\mathbf{u}(t)+\mathbf{b}\; \xi(t),
 			\end{equation}
 			where $\xi(t)$ is white noise with  $E\{\xi(t)\}=0$ and $E\{\xi(t)\xi(t+\tau)\}=\sigma^2\delta(\tau)$,
 			$\;\sigma\in \R$,  $\delta(\cdot)$ is the Dirac function,
 			\begin{equation}
 			\begin{aligned}\label{Zustandsmatrix}
 			\mathrm{\mathbf{A}}=\left(\begin{matrix}
 			-a_1&  1 & 0 & \cdots & 0\\
 			-a_2&  0 & 1 & \ddots & \vdots\\
 			\vdots& \vdots &\ddots&\ddots& 0\\
 			-a_{p-1}& 0 &   \cdots   &0&1 \\
 			-a_p& 0    &\cdots    &\cdots  &0
 			\end{matrix}\right),\;\;\;
 			\mathbf{b}=\left(\begin{array}{c}
 			b_{p-1}\\
 			b_{p-2}\\
 			\vdots\\
 			b_1\\
 			b_0\\
 			\end{array}\right),\;\;\;
 			\mathbf{c}= \left(\begin{array}{c}
 			1\\
 			0\\
 			\vdots\\
 			0\\
 			\end{array}\right),
 			\end{aligned}
 			\end{equation}
 			and $b_j=0$ if $j>q$.
 		\end{definition}
 		%
 		%
 		%
 		Equation \eqref{def:CARMApqZustandsformDGL} can be expressed as the stochastic differential equation
 		\begin{equation}\label{eq:SDE_CARMApq}
 		\mathrm{d}\mathbf{u}(t)=\mathrm{\mathbf{A}}\,\mathbf{u}(t) \,\mathrm{d}t+\mathbf{b}\, \mathrm{d}W(t),
 		\end{equation}
 		using the relation $\mathrm{d}W(t)=\xi(t)\, \mathrm{d}t $ between the white noise $\xi(t)$ and the  increment of the Wiener process $\mathrm{d}W(t)$.
 		The above defined CARMA(p,q) process has the transfer function
 		\begin{equation}\label{eq:TransferCARMApq}
 		H_{\mathrm{carma}}(s)=\frac{b_0+b_1s+\ldots+b_qs^q}{s^p+a_1 s^{p-1}+\ldots+a_p},
 		\end{equation}
 		where $s:=\mathrm{i}\omega$,
 		and the spectral density
 		\begin{equation}\label{eq:LapCARMApq}
 		S_{\mathrm{carma}}(\omega)=\sigma^2\frac{|b_0+b_1s+\ldots+b_qs^q|^2}{|s^p+a_1 s^{p-1}+\ldots+a_p|^2}.
 		\end{equation}
 		Thereby, $\omega$ is the angular frequency.
 		It should be noted, that the state space representation in Definition \ref{def:CARMApqZustandsform} is not unique.
 		%




\end{document}